\documentclass[nofootinbib,longbibliography,prd,floatfix,superscriptaddress,twocolumn]{revtex4-1}
\usepackage[english]{babel}
\usepackage{amsmath,amssymb,amsbsy,amstext,amsthm,booktabs,simplewick,exscale,relsize,graphicx,amsfonts,
upgreek,slashed,color,multirow,dcolumn,bm,enumerate,url,hyperref,chngcntr,xcolor}

\begin{document}

\title{Searching for Dark Matter with Neutron Star Mergers and Quiet Kilonovae}

\author{Joseph Bramante}
\affiliation{CPARC and Department of Physics, Engineering Physics, and Astronomy, Queen's University,  Kingston, Ontario, Canada}
\affiliation{Perimeter Institute for Theoretical Physics, Waterloo, Ontario, Canada}

\author{Tim Linden}
\affiliation{CCAPP and Department of Physics, The Ohio State University, Columbus, OH, USA}

\author{Yu-Dai Tsai}
\affiliation{Perimeter Institute for Theoretical Physics, Waterloo, Ontario, Canada}
\affiliation{Laboratory for Elementary Particle Physics, Cornell University, Ithaca, NY, USA}

\begin{abstract}
We identify new astrophysical signatures of dark matter that implodes neutron stars (NSs), which could decisively test whether NS-imploding dark matter is responsible for missing pulsars in the Milky Way galactic center, the source of some $r$-process elements, and the origin of fast-radio bursts. First, NS-imploding dark matter forms $\sim  10^{-10}$ solar mass or smaller black holes inside neutron stars, which proceed to convert neutron stars into $\sim$1.5 solar mass BHs. This decreases the number of neutron star mergers seen by LIGO/Virgo (LV) and associated merger kilonovae seen by telescopes like DES, BlackGEM, and ZTF, and instead, producing a population of ``black mergers" containing $\sim$1.5 solar mass black holes. Second, dark matter-induced neutron star implosions may create a new kind of kilonovae that lacks a detectable, accompanying gravitational signal, which we call ``quiet kilonovae." Using DES data and the Milky Way's r-process abundance, we constrain quiet kilonovae. Third, the spatial distribution of neutron star merger kilonovae and quiet kilonovae in galaxies can be used to detect dark matter. NS-imploding dark matter destroys most neutron stars at the centers of disc galaxies, so that neutron star merger kilonovae would appear mostly in a donut at large radii. We find that as few as ten neutron star merger kilonova events, located to $\sim$1~kpc precision could validate or exclude dark matter-induced neutron star implosions at $2 \sigma$ confidence, exploring dark matter-nucleon cross-sections 4-10 orders of magnitude below current direct detection experimental limits. Similarly, NS-imploding dark matter as the source of fast radio bursts can be tested at $2 \sigma$ confidence once 20 bursts are located in host galaxies by radio arrays like CHIME and HIRAX. 
\end{abstract}

\maketitle
\section{Introduction}
\label{sec:intro}

Uncovering the identity and interactions of dark matter would deepen our understanding of fundamental physics and the origin of our universe. Because dark matter is abundant in galactic halos, it may be revealed through its impact on stars. Indeed, a number of ongoing astrophysical searches may be unveiling dark matter that forms black holes inside and implodes old neutron stars \cite{Bramante:2014zca,Bramante:2015dfa,Goldman:1989nd,deLavallaz:2010wp,Kouvaris:2010jy}, thereby creating r-process elements \cite{Bramante:2016mzo}, and possibly fast radio bursts \cite{Fuller:2014rza}. 

Traditional neutron star-based searches for asymmetric dark matter have relied on old pulsars in the Milky Way, whose old age bounds the capture rate of asymmetric dark matter, which can eventually form black holes inside of neutron stars. Finding old pulsars in the Milky Way's galactic center, where dark matter is denser, more dark matter is captured, and therefore pulsars potentially implode faster, would advance the frontier of asymmetric dark matter detection. However, extensive radio surveys of the Milky Way galactic center have not found enough pulsars to conclusively strengthen or invalidate the hypothesis that dark matter implodes neutron stars in that region.
Fortuitously, the unprecedented sensitivity of laser interferometer gravitational wave detectors at LIGO/Virgo \cite{Abbott:2016ymx}, the broad optical purview of DES \cite{Doctor:2016gdi,Soares-Santos:2016qeb}, BlackGEM \cite{Ghosh:2014yga}, and other optical telescopes, as well as kilo-channel radio reception at CHIME \cite{Berger:2016ejd} and HIRAX \cite{Newburgh:2016mwi}, are better scrutinizing the dynamics of neutron stars and black holes. Indeed, recently a neutron star merger was found both in gravitational waves \cite{TheLIGOScientific:2017qsa} and follow-on telescope observations \cite{GBM:2017lvd}.

This article demonstrates how the combined statistics of neutron star mergers observed in galaxies can be used to unmask dark matter with nucleon scattering cross-sections orders of magnitude smaller than the present reach obtained from neutron stars in the Milky Way, and up to ten orders of magnitude beyond any planned underground dark matter detection experiment. This new search method applies to many asymmetric dark matter models \cite{Petraki:2013wwa,Zurek:2013wia}, including $\rm keV-PeV$ mass bosons \cite{Bramante:2014zca,Bramante:2015dfa,Goldman:1989nd,deLavallaz:2010wp,Kouvaris:2010jy,McDermott:2011jp,Bramante:2013hn,Bell:2013xk,
Bertoni:2013bsa,Guver:2012ba,Kouvaris:2013kra,Kurita:2015vga}, $\rm keV-EeV$ mass self-interacting fermions \cite{Kouvaris:2011gb,Bramante:2013nma,Bramante:2015dfa}, and $\rm \gtrsim PeV$ mass bosons and fermions \cite{deLavallaz:2010wp,Bramante:2015cua,Bramante:2016mzo,Bramante:2017obj}. 

Additionally, this article details distinct signatures of dark matter-induced neutron star implosions, which can be discovered by upcoming gravitational, optical, and radio surveys. NS-imploding dark matter may produce detectable quiet kilonovae: a kilonova powered by the ejection of neutron star fluid during the dark matter-induced implosion of a neutron star. These \emph{quiet kilonovae} would be ``quiet" and distinct from the now-observed neutron star \emph{merger kilonovae}, in that they would not produce gravitational waves detectable by LIGO/Virgo. The remainder of this document is structured as follows: in Section \ref{sec:dmNSimp} we review asymmetric dark matter that implodes neutron stars and introduce the normalized implosion time. Section \ref{sec:pop} details how neutron star populations in disc galaxies are altered by NS-imploding dark matter. Section \ref{sec:nspbh} details prospects for finding primordial black holes (PBHs) using neutron star mergers and quiet kilonova.  Section \ref{sec:qk} provides the first constraint on ``quiet kilonovae." Section \ref{sec:dmsearch} details a major result of this work: that NS mergers can be used as an incisive new search for dark matter. A location and rate analysis testing whether fast radio bursts originate from dark matter-induced neutron star implosions is presented in Section \ref{sec:frbs}. In section \ref{sec:conc}, we conclude.  Appendix \ref{app:cdfs} presents the cumulative distributions functions employed in Section \ref{sec:frbs} and \ref{sec:dmsearch} and Appendix \ref{app:ADM} provides details of neutron star implosions induced by heavy asymmetric dark matter. 

\section{Dark matter-induced neutron star implosions}  
\label{sec:dmNSimp}
Once enough dark matter has accumulated in a neutron star's interior, dark matter may collapse into a small ($\lesssim 10^{-10}~ {\rm M_{ \odot}}$) black hole that subsequently consumes the neutron star \cite{Bramante:2014zca,Goldman:1989nd,Bramante:2016mzo}. We begin by defining a useful variable combination, the ``normalized implosion time," which relates dark matter-induced NS implosions occurring at different radii from a galactic center, where the local dark matter density $\rho_{\rm x}$ and velocity dispersion $v_{\rm x}$ will be different, see Figure \ref{fig:adonuts}. The maximum mass accumulation rate of dark matter into a NS is \cite{Baryakhtar:2017dbj}
\begin{align}
 \dot{m}_{\rm x} &= \pi \rho_{\rm x} \frac{2G M R}{v_{\rm x}} \left(1- \frac{2 G M}{R} \right)^{-1} \nonumber \\ &\simeq {\rm \frac{10^{26}~GeV}{s}} \left(\frac{\rho_{\rm x}}{\rm GeV/cm^3} \right) \left(\frac{\rm 200~km/s}{v_{\rm x}} \right),
\end{align} 
where $M$ and $R$ are the mass and radius of the neutron star and $G$ is Newton's constant. The time until NS implosion scales inversely with the mass accumulation rate, $t_{\rm c} \propto \dot{m}_{\rm x}^{-1}$; therefore $t_{\rm c}$ is proportional to the dark matter velocity dispersion divided by density, $t_{\rm c} \propto  v_{\rm x} / \rho_{\rm x} $. Furthermore, $v_{\rm x}$  and $\rho_{x}$ are the only quantities in $\dot{m}_{\rm x}$ that depend on the galactocentric radius $r$. It follows that for dark matter which implodes NSs in time $t_{\rm c}$, the quantity
\begin{align}
t_{\rm c} \frac{\rho_{\rm x}}{v_{\rm x} } = {~\rm Constant} \times \left[{\rm Gyr} ~\frac{\rm GeV/cm^{3}}{\rm 200~ km/s} \right],
\label{eq:defradinv}
\end{align}
which we call the normalized implosion time,\footnote{The units given in square brackets in  Eq.\eqref{eq:defradinv} might be read as ``A neutron star will implode in one gigayear for local dark matter density ${\rm GeV/cm^3}$ and dark matter velocity dispersion $200 \rm ~km/s$."} is independent of $r$. Throughout we will normalize $t_{\rm c} \rho_{\rm x} /v_{\rm x} $ to a typical dark matter density ($\rm GeV/cm^3$) and velocity dispersion ($\rm 200~km/s$) for a disc galaxy. 

The value of $t_{\rm c} \rho_{\rm x} /v_{\rm x} $ for a specific dark matter model can be determined by calculating the time for dark matter with local density $\rho_{\rm x}$ and relative velocity $v_{\rm x}$ to implode a neutron star. While many asymmetric dark matter models implode neutron stars \cite{Bramante:2014zca,Bramante:2015dfa,Goldman:1989nd,deLavallaz:2010wp,Kouvaris:2010jy,McDermott:2011jp,Bramante:2013hn,Bell:2013xk,
Kouvaris:2011gb,Bramante:2013nma,Bramante:2015dfa,Bramante:2015cua,Bramante:2016mzo}, we will focus on heavy $m_{\rm x} \gtrsim {\rm PeV}$ asymmetric dark matter as a simple example. In the case of heavy asymmetric dark matter, the critical mass of dark matter required to form a small black hole is $M_{\rm crit}^{\rm f} \sim \frac{m_{\rm pl}^3}{m_{\rm x}^2}$ for dark fermions with mass $m_{\rm x}$ \cite{Bramante:2015cua}, and $M_{\rm crit}^{\rm b} \sim 0.12 \sqrt{\lambda} \frac{m_{\rm pl}^3}{m_{\rm x}^2}$ for dark scalars with self-interaction potential $V(\phi) = \lambda |\phi|^4$ \cite{Colpi:1986ye}. In these models, the neutron star will implode shortly after it collects a critical mass of dark matter at time $t_{\rm c} \simeq M_{\rm crit} / \dot{m}_{\rm x}$, where this expression assumes all dark matter passing through the neutron star is captured -- see Appendix \ref{app:ADM} for details and for the scaling of dark matter-nucleon cross-section with $t_{\rm c} \rho_{\rm x} /v_{\rm x} $. Then the value of the galactic radial invariant $t_{\rm c} \rho_{\rm x} /v_{\rm x} $  is
\begin{align}
t_{\rm c} \frac{\rho_{\rm x}}{ v_{\rm x} } \Big|_{\rm f} &= \left( \frac{\rm 10~PeV}{m_{\rm x}} \right)^2~15~ {\rm Gyr}~ \frac{\rm GeV/cm^{3}}{\rm 200~ km/s} 
\nonumber \\
t_{\rm c} \frac{\rho_{\rm x}}{ v_{\rm x} } \Big|_{\rm b} &= \left( \frac{\lambda}{1} \right)^{1/2} \left( \frac{\rm 3~PeV}{m_{\rm x}} \right)^2~20~ {\rm Gyr} ~\frac{\rm GeV/cm^{3}}{\rm 200~ km/s},
\label{eq:tc}
\end{align}
for heavy asymmetric fermions and bosons, respectively.

\section{Black mergers, quiet kilonovae, and r-process donuts}
\label{sec:pop}
NS-imploding dark matter creates an unexpected population of low mass $\sim1.5~{\rm M_{\odot}}$ black holes (BHs), depleting the expected population of NSs. This in turn would alter the number of merging neutron stars that would be seen by LIGO/Virgo, along with their accompanying merger kilonovae, which are the days-long luminous outbursts from beta decaying neutrons ejected when NSs fall into a BH or another NS \cite{1977ApJ...213..225L,Kasen:2013xka}.

\begin{figure}[t]
\includegraphics[width=0.45\textwidth]{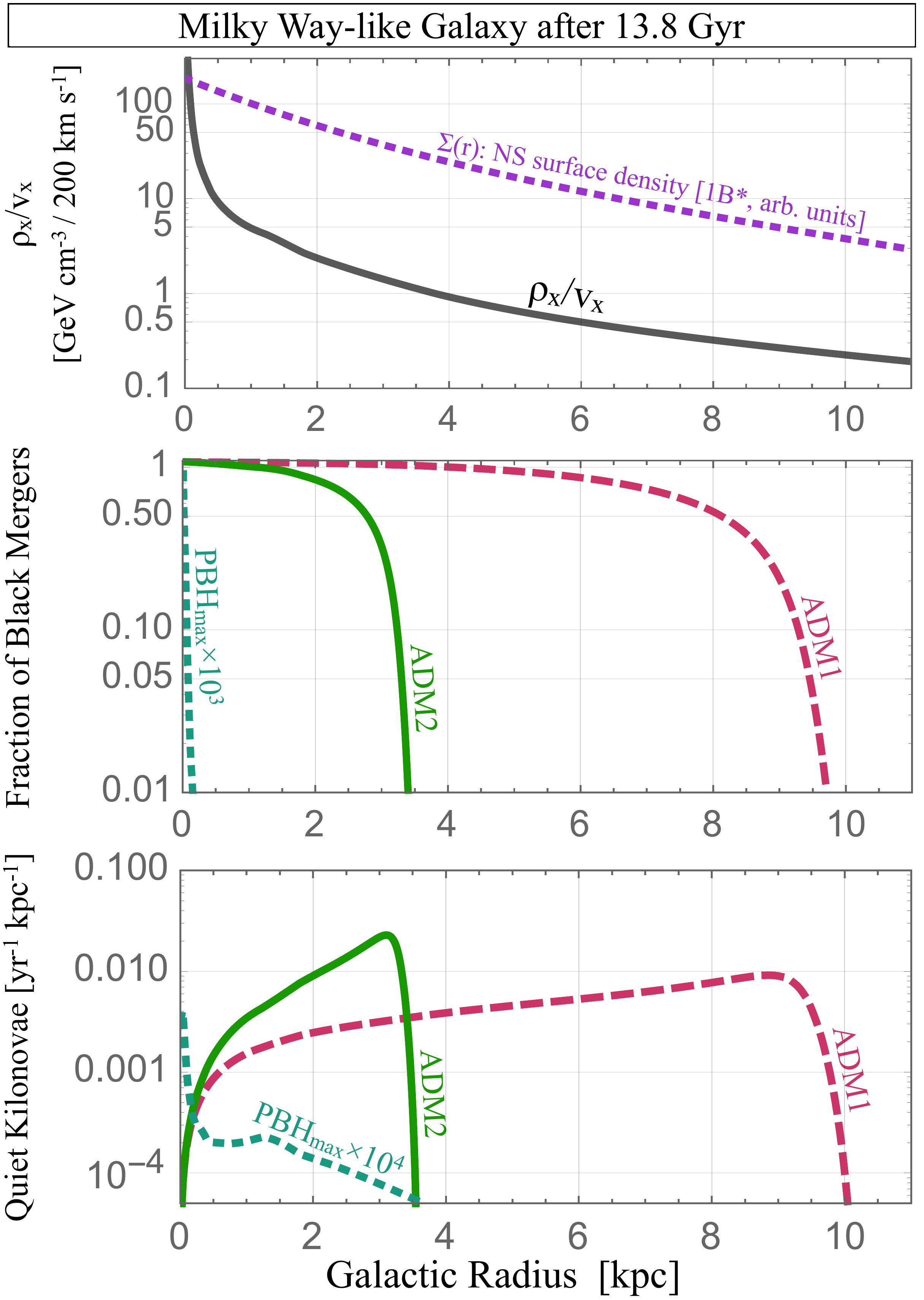}
\caption{The dark matter density over velocity $\rho_{\rm x}/v_{\rm x}$ and NS surface density $\Sigma$ in an MWEG (top), the total fraction of imploded neutron stars (middle), and the rate for quiet kilonovae, aka dark matter-induced NS implosions, per yr per kpc (bottom), are given as a function of distance from the center of a Milky Way-like galaxy for two asymmetric dark matter models ADM1 and ADM2, defined by $t_{\rm c} \rho_{\rm x} /v_{\rm x}  = 3$ and $15$ $\rm Gyr~GeV/cm^3~ (200 ~km/s)^{-1}$. PBH$_{\rm max}$ is a maximally NS-imploding model described in Appendix \ref{app:ADM}. For visibility, the PBH curve has been augmented by 3 and 4 orders of magnitude. }
\label{fig:adonuts}
\end{figure}

We now determine the number and position of neutron stars converted to BHs by dark matter in a Milky Way-like galaxy. With some subtleties that we will address, our findings for a typical 13 Gyr old $\sim 10^{12} \rm ~M_{\odot}$ disc galaxy can be applied to events in different galaxies, using a Milky Way equivalent galaxy (MWEG) volumetric conversion for merger and kilonova rates, $i.e.$ one ${\rm MWEG}~{\rm per} ~(4.4~{\rm Mpc})^3$ \cite{Abadie:2010cf}. 

The number of neutron stars converted into BHs by dark matter in an MWEG will depend on the historic neutron star formation rate in the galaxy, the dynamics and final positions of neutron stars after formation, the dark matter halo density profile, and the relative velocity of dark matter with respect to neutron stars. We model the historic star formation rate $\dot{M^*}(t)$ using a global fit to astronomical data (\cite{Hopkins:2006bw}, Table 1, Column 2). While we use $\dot{M}^*(t)$ to determine the relative historic rate of neutron star formation, we normalize the total rate to $10^9$ neutron star births over the MWEG lifetime \cite{1990ApJ...348..485P,1993ApJ...403..690B,2010A&A...510A..23S}. At birth, it has been observed that neutron stars receive natal kicks which result in an initial velocity boost of $\sim 250 ~{\rm km/s}$~\citep{Hobbs:2005yx,FaucherGiguere:2005ny}. A recent study of neutron star dynamics in an MWEG has found that most ($\gtrsim 80 \%$) neutron stars are retained within a $\sim$kiloparsec of the MWEG disc plane, with a NS surface density $\Sigma(r)$. We therefore model the MWEG neutron star distribution as a thin disk with surface density $\Sigma(r)$ given by models 1B* (and 1C* as indicated for comparison) in \cite{2010A&A...510A..23S}.

To model dark matter in an MWEG, we use an NFW dark matter halo density profile \cite{Navarro:1996gj}, \mbox{$\rho_{\rm x}(r) = \rho_0 (r/R_{\rm s})^{-1}(1+r/R_{\rm s})^{-2}$}, with dark matter density normalization $\rho_0 = 0.3 ~{\rm GeV/cm^3}$ and scale factor $R_{\rm s}=20~{\rm kpc}$. To approximate the dark matter velocity dispersion in an MWEG, we match the phenomenological fit of Sofue to stellar velocities in the Milky Way (\cite{Sofue:2013kja}, Figure 11).  With the star formation rate, neutron star distribution and dark matter properties specified, the fraction of neutron stars at radius $r$ converted to solar mass BHs is given by
$
F_{\rm BH}(r) = \frac{ \int_0^{{\rm Max}[t_{\rm u} - t_{\rm c} (r),0]} \dot{M^*}(t)~dt}{ \int_0^{t_{\rm u}} \dot{M^*}(t)~dt},
$
where $t_{\rm u} \sim 13.8 ~{\rm Gyr}$ is the lifetime of the universe and $t_{\rm c}(r)$ is the collapse time at radius $r$, obtainable from Eq.~\eqref{eq:tc}. Similarly, the rate of neutron star implosions (and also quiet kilonovae) per unit galactocentric radius is given by
$
R_{\rm qk} = 2 \pi r \Sigma(r) \dot{M}^*(t_{\rm u} -  t_{\rm c} (r)).
$
In Figure \ref{fig:adonuts}, we plot the fraction of neutron stars converted to BHs along with the rate of neutron star implosions per year per kpc, both as a function of galactocentric radius, for a 13 Gyr old MWEG. In Table \ref{tab:rates}, we show how standard rates for compact object mergers would be altered, and display dark matter-induced neutron star implosion rates, for a few values of $t_{\rm c} \rho_{\rm x} /v_{\rm x} $. Table \ref{tab:rates} also gives the maximum rate for PBH implosion of NSs, which we address in the next section.
\begin{table}[h]
\begin{center}
\begin{tabular}{| c | c | c | c | c | c | c |}
\hline
Model &NS-NS&NS-BH&BH-BH&LM-BH&NS Im.&Im./$t_{\rm u}$\\ \hline \hline
Non-Imp. &1e-4&3e-6&4e-7&0&0&0 \\ \hline
ADM1 &3e-5&9e-7&4e-7&7e-5&4e-2&7e8\\ \hline
ADM2 &7e-5&2e-6&4e-7&3e-5&3e-2&3e8\\ \hline
PBH$_{\rm max}$ &1e-4&3e-6&4e-7&4e-11&1e-7&400   \\ \hline
\end{tabular}
\end{center}
\caption{The first five columns give the rate for compact object mergers and dark matter-induced neutron star implosions per MWEG per year ($AeB \equiv A \times 10^{B}$), for both ``Non-Implosive" and NS-imploding dark matter. ADM1 and ADM2 are defined by $t_{\rm c} \rho_{\rm x} /v_{\rm x}  = 3$ and $15$ $\rm Gyr~GeV/cm^3~ (200 ~km/s)^{-1}$ respectively, and PBH$_{\rm max}$ is a maximally NS-imploding primordial BH model defined in Section \ref{sec:nspbh}. NS-NS, NS-BH, and BH-BH indicate standard NS and BH mergers, while LM-BH indicates a BH-BH merger with at least one $\sim 1.5~M_{\odot}$ BH (Black Merger). We use the average BH and NS merger rates predicted in \cite{Abadie:2010cf}; actual merger rates may be 100-fold larger or smaller. The final column shows the number of NS implosions expected in a $t_{\rm u} \sim 13$ Gyr old MWEG hosting $10^9$ 1B*-distributed NSs.}
\label{tab:rates}
\end{table}

\section{Rare neutron star implosions from primordial black holes} 
\label{sec:nspbh}
Black holes formed from primordial perturbations during the radiation-dominated expansion of the early universe \cite{Hawking:1974rv,Carr:1974nx}, with masses between $\sim 10^{41} - 10^{50}$ GeV, can be captured inside and subsequently consume neutron stars \cite{Capela:2013yf,Capela:2012jz}. As this work was being completed \cite{Fuller:2017uyd} appeared, which addresses PBH-induced NS implosions, and following \cite{Bramante:2016mzo}, considers r-process elements and kilonovae produced by NS implosions. The maximum PBH-induced NS implosion rate for an MWEG found here differs markedly from \cite{Fuller:2017uyd}, because we use the realistic, standard values for the NS population density, PBH density, and PBH velocity dispersion. We will find that NS implosions from primordial BHs (PBHs) in a typical Milky Way-like galaxy are rare. 

PBHs with halo density $\rho_{\rm pbh}$ are captured by neutron stars at a rate \cite{Capela:2013yf}
\begin{align}
C_{\rm pbh}=\sqrt{6\pi}\frac{\rho_{\rm pbh}}{m_{\rm pbh}}\left(\frac{2 G M R}{v_{\rm x}}\right) \frac{1-{\rm Exp}\left[-\frac{3E_{\rm loss}}{m_{\rm pbh}v_{\rm x}^2}\right]}{1-\frac{2GM}{R}},
\label{eq:pbhcap}
\end{align}
where the energy loss of a PBH transiting the NS is $E_{\rm loss} \simeq \frac{4 G^2 m^2_{\rm pbh}M}{R^2} \left\langle  \frac{\ln\Lambda}{2 G M / R} \right\rangle$, and for a typical neutron star density profile $ \left\langle \frac{\ln\Lambda}{2 G M / R}\right\rangle \sim 14.7$. With Eq.~\eqref{eq:pbhcap} it can be verified that PBH capture in NSs is maximized for PBH masses $m_{\rm pbh} \sim 10^{44}-10^{47}$ GeV. Assuming $m_{\rm pbh} \sim 10^{45}$ GeV PBHs make up the entire dark matter density, $\rho_{\rm pbh} \simeq \rho_{\rm x}$, we find that the PBH NS implosion rate appears too low to be detectable by next generation astronomical surveys, as shown in Figure 1\ref{fig:adonuts} and Table \ref{tab:rates}.

\section{Milky Way r-process enrichment and DES bounds on quiet kilonovae}
\label{sec:qk}
\begin{figure}[t!]
\includegraphics[width=0.45\textwidth]{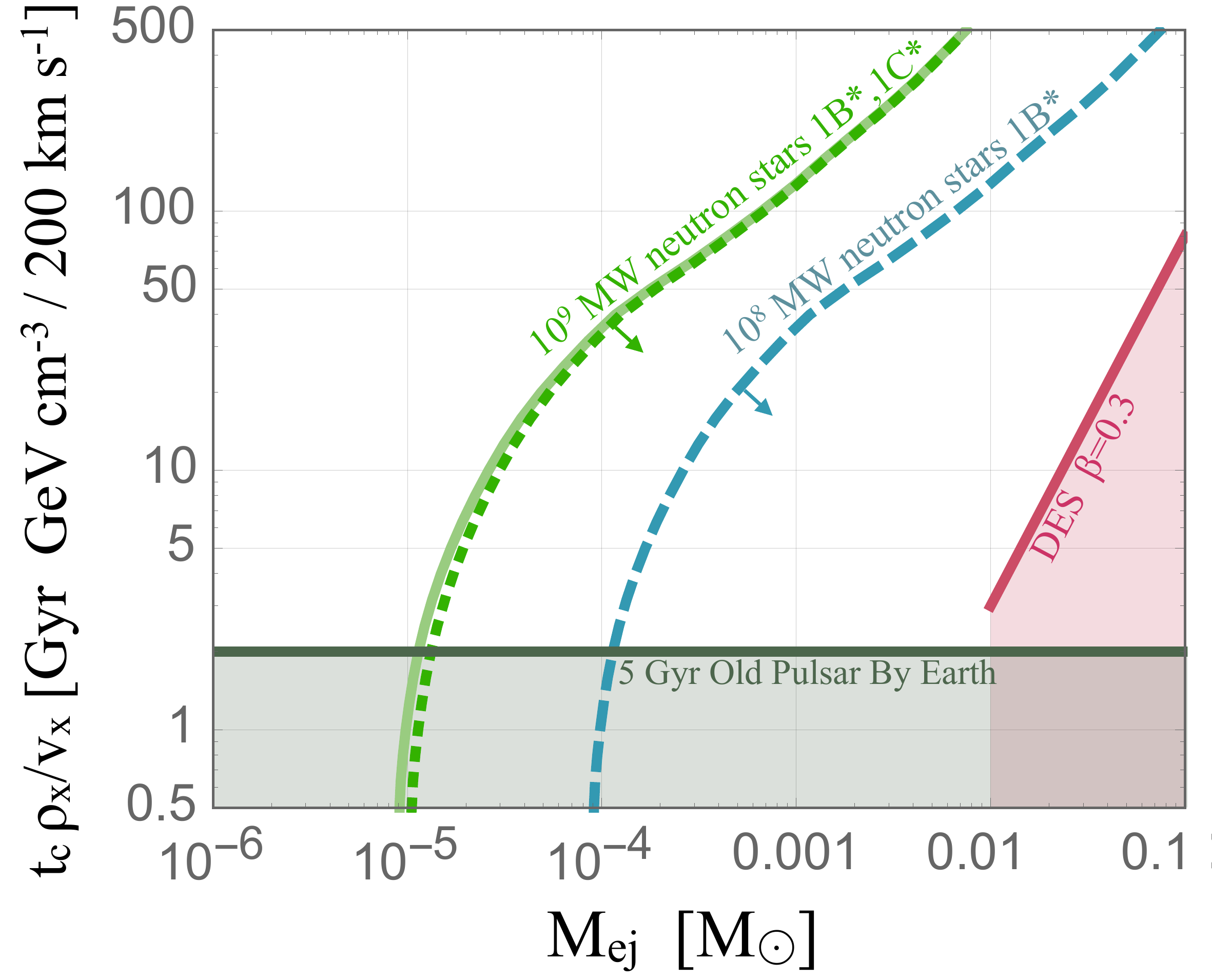}
\caption{Bounds on dark matter-induced quiet kilonovae as a function of the neutron fluid mass ejected $M_{\rm ej}$ during a NS implosion. The DES bounds assume kilonovae with an ejection velocity $\beta=0.3$c \cite{Doctor:2016gdi}, and assume $10^9$ NSs in an MWEG. Milky Way $r$-process elements produced from $M_{\rm ej}$ per NS implosion imply the indicated bounds \cite{Bramante:2016mzo}, for either $10^8$ or $10^9$ total NSs, assumed to have spatial distribution 1B* or 1C* given in \cite{2010A&A...510A..23S} (arrows indicate the direction of excluded regions). Old pulsars in the Milky Way \cite{Bramante:2014zca,Bramante:2015dfa} also set constraints. The x-axis indicates dark matter that implodes NSs in time $t_{\rm c}$ for background dark matter density $\rho_{\rm x} $ with velocity dispersion $v_{\rm x}$, expressed in units of $ t_{\rm c} \rho_{\rm x} / v_{\rm x}$, see Eq.~\eqref{eq:defradinv}.}
\label{fig:mejecta}
\end{figure}
NSs imploded by dark matter may eject a substantial amount of neutrons into the interstellar medium. Ejected neutron fluid will decompress, beta decay, and form a portion of the r-process elements observed in the Milky Way \cite{1977ApJ...213..225L,1989Natur.340..126E,Kasen:2013xka,Bramante:2016mzo}. R-process elements are heavy elements with atomic masses around $80, 130,195$, formed from neutron rich fluid at an as-yet undetermined astrophysical site. While core collapse supernovae have been historically favored as candidate sites for r-process production, recent observations of a high r-process abundance in Reticulum II, and low r-process abundance in other ultra faint dwarf galaxies, favors r-process production from rare events like a NS merger \cite{2016Natur.531..610J} or NS implosion \cite{Bramante:2016mzo}. In the case of a NS implosion, the amount of NS fluid ejected will likely depend on tidal forces during the implosion \cite{Bramante:2016mzo}, which require a complete hydrodynamical simulation to be properly modelled.  However, it is known that in total,  $\sim 10^4~M_{\odot}$ of r-process elements must be formed to match the abundance seen in the Milky Way \cite{2015MNRAS.447..140V,2015ApJ...807..115S,Wehmeyer:2015sra}. Therefore, the amount of neutron fluid ejected per NS implosion can be bounded, by limiting the total NS mass ejected to $\sim 10^4~M_{\odot}$ in the Milky Way. In Figure \ref{fig:mejecta}, we present such bounds, as a function of $t_{\rm c} \rho_{\rm x} /v_{\rm x}$. This can be compared to the final column of Table \ref{tab:rates}, which shows the expected number of NS implosions after $\sim 13$ Gyr.

Quiet kilonovae produced by NS-imploding dark matter can be searched for using state-of-the-art optical surveys. DES has recently published a null wide field optical search for kilonovae \cite{Doctor:2016gdi}, which are the days-long luminous outbursts of beta-decaying neutron fluid ejected from NSs falling into BHs or other neutron stars. Because this search does not rely on a gravitational signature and instead seeks out beta decay emission from NS fluid flung into outer space, its findings can be used to constrain quiet kilonovae, $i.e.$ NS fluid ejected from a NS implosion. Because kilonovae light curves depend mainly on the mass and velocity of NS fluid ejected \cite{Kasen:2013xka}, bounds obtained for NS merger kilonovae can be applied to quiet kilonovae from NS implosions. We set this bound in Figure \ref{fig:mejecta}, computing the quiet kilonova rate for each $ t_{\rm c} \rho_{\rm x} / v_{\rm x}$ model point, assuming an MWEG containing $10^9$ NSs.

\begin{figure}[h!]
\includegraphics[width=0.49\textwidth]{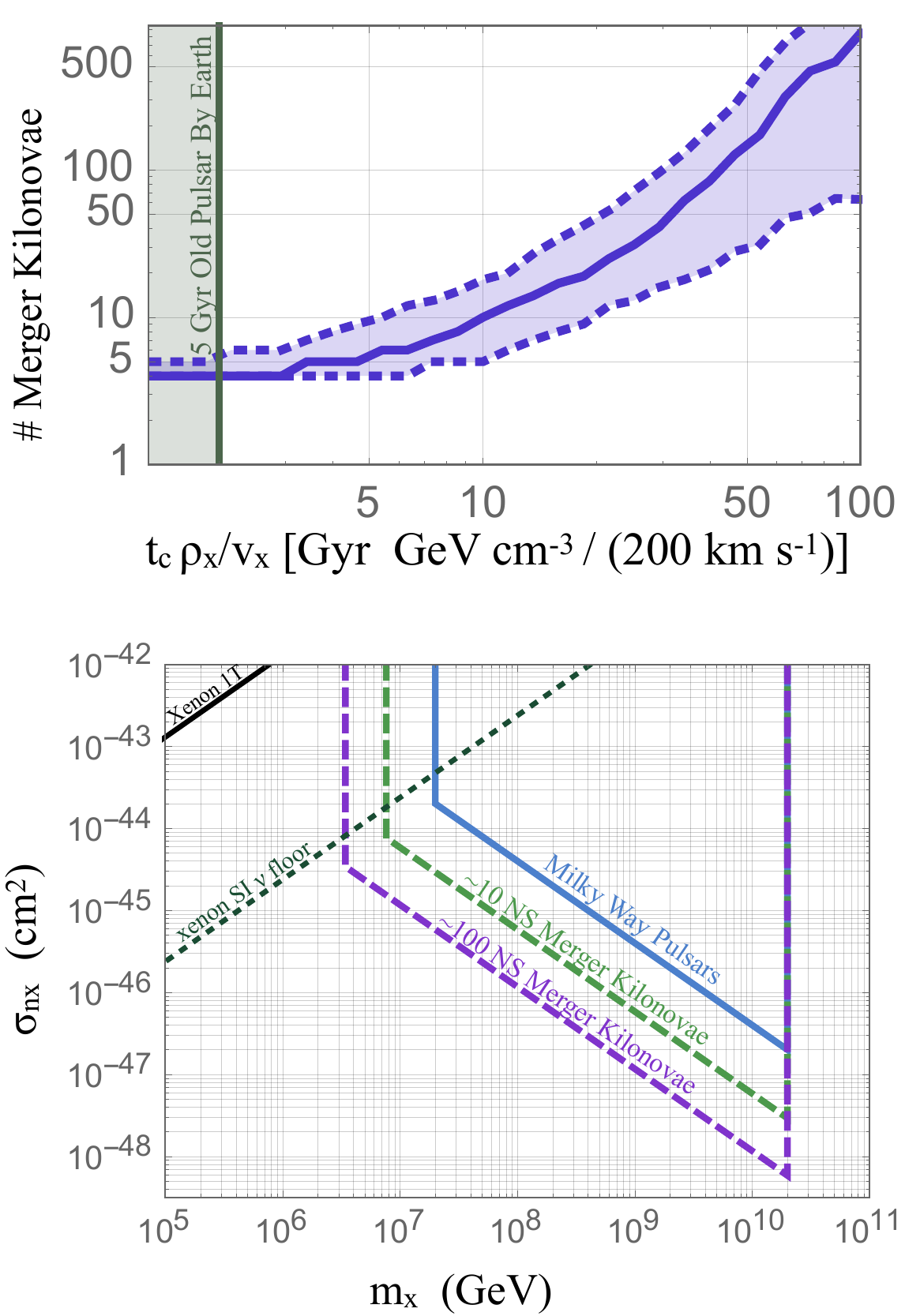}
\caption{(Top) The number of NS mergers found by LIGO/Virgo, located to within $\sim 1 {~ \rm kpc}$ in a host galaxy by optical imaging of a kilonova, required to exclude dark matter that implodes NSs in time $t_{\rm c}$ for background dark matter density $\rho_{\rm x} $ with velocity dispersion $v_{\rm x}$, expressed in units of $t_{\rm c}  \rho_{\rm x} / v_{\rm x}$, see Eq.~1 in the main text. A Kolmogorov-Smirnov test was performed at each model point against the standard hypothesis that NS mergers track a standard distribution of NSs (see Ref.~\cite{2010A&A...510A..23S}, model 1B*) in a disc galaxy, for four hundred randomly generated merger data sets, allowing for up to $10^3$ NS merger events per data set. Dotted bands indicate number of events needed for $2 \sigma$ significance, for the upper and lower quartile of randomly generated merger data sets; the solid line indicates the median. 
(Bottom) The fermionic asymmetric dark matter-nucleon cross-section sensitivity obtainable with future observation of $\sim 10$ and $\sim 100$ NS merger kilonovae is shown, along with the MW pulsar bound \cite{Bramante:2015dfa}, Xenon 1T bound \cite{Aprile:2017iyp}, and the xenon neutrino floor \cite{Ruppin:2014bra}, below which atmospheric neutrinos provide an irreducible background to dark matter scattering. Note that for fermionic dark matter masses $\gtrsim 10^{10}$ GeV, the black holes formed in  NSs are too small and evaporate via Hawking radiation \cite{Bramante:2013hn}.}
\label{fig:loudonut}
\end{figure}

\section{Searching for dark matter with NS mergers}
\label{sec:dmsearch}
Here we show how the galactocentric radial positions of $\sim 10$ merger kilonovae would be sufficient to explore asymmetric dark matter-nucleon cross-sections orders of magnitude smaller than those presently probed using old pulsars in the Milky Way. The current generation of LV instrumentation is sensitive to gravitational strains on the order of 10$^{-23}$ at an optimal frequency of 400~Hz, allowing for the observation of double neutron star (NS) binaries out to distances of $\sim$70~Mpc \cite{Abbott:2016ymx}. Anticipated upgrades will significantly expand this reach, as the amplitude of gravitational wave events is inversely proportional to the source distance, while the expected merger rate increases as the distance cubed. In the coming decade, up to hundreds of NS merger events are anticipated. Once a NS merger event is located to within $\sim 10$ square degrees by LIGO/Virgo, wide field telescopes like BlackGEM \cite{Ghosh:2014yga} and the Zwicky Transient Factory \cite{2014htu..conf...27B} are poised to image any subsequent kilonovae. The number of kilonovae found using this method will depend on their peak brightness, predictions for which range from -10 to -16 AB magnitude \cite{Doctor:2016gdi}, while $e.g.$ BlackGEM will perhaps probe a 100 degree field of view down to -14 AB magnitude for 200 Mpc distant mergers. As shown in Figure \ref{fig:adonuts}, galaxies with NS-imploding dark matter will have fewer NS merger kilonovae in their centers, where most NSs will have been converted to BHs. Therefore, the spatial distribution of merger kilonovae that can be used to test for dark matter. In the top panel of Figure \ref{fig:loudonut}, we show the results of a cumulative distribution test, where the standard (model ``1B*") NS distribution defined Section 2, is tested against the distribution expected if dark matter is imploding neutron stars. 

The altered NS merger distribution is calculated by taking the fraction of NSs converted into black holes shown in Figure \ref{fig:adonuts}, and applying this conversion fraction to the 1B* expected distribution of NS mergers. The expected and dark matter-modified cumulative distribution functions of NS mergers in an MWEG are plotted in Appendix \ref{app:cdfs}. Statistical results were obtained by running 400 random Kolomogorov-Smirnov cumulative distribution trials, for each neutron star normalized implosion time ($t_{\rm c} \rho_{\rm x} /v_{\rm x}$), to determine how many merger kilonovae located in galaxies would be necessary to detect NS-imploding dark matter at 2$\sigma$ significance. Using the same methodology, in Section \ref{sec:frbs} we find that $\sim 20$ FRBs localized in galaxies would determine whether FRBs are a byproduct of NS implosions. 

In practice, merger kilonovae occur in galaxies that are somewhat different from the Milky Way. To convert a measured galactocentric radius in a NS-merger-containing (non-Milky Way) galaxy, $r_{\rm nMW}$, to a Milky Way equivalent radius $r_{\rm MW}$, one can solve the formula
\begin{align}
\frac{\rho_{\rm x}^{\rm MW}({r_{\rm MW}})}{v_{x \rm}^{\rm MW}({r_{\rm MW}})} = \frac{\rho_{\rm x}^{\rm nMW}({r_{\rm nMW}})}{v_{x \rm}^{\rm nMW}({r_{\rm nMW}})}
\label{eq:convert}
\end{align}
for $r_{\rm MW}$, where $\rho_{\rm x}$ and $v_{\rm x}$ are the dark matter density and velocity dispersion of the MWEG and non-Milky Way galaxies, as indicated. For example, the recently detected NS merger in NGC 4993 occurred in a $\sim 10^{10.9} ~M_{\rm \odot}$ galaxy which would have an NFW profile defined by $\rho_0 = 0.34 ~{\rm GeV/cm^3}$ and scale factor $R_{\rm s}=7.5~{\rm kpc}$. The NS merger in NGC 4993 occurred at$ \sim 2-3$ kpc from its center \cite{GBM:2017lvd}. Solving Eq.~\eqref{eq:convert}, this corresponds to an Milky Way equivalent radius of 5-8 kpc. Note that this analysis also assumes that most identified NS mergers will have an age of $\sim 10$ Gyr -- indeed, the NS merger progenitor found in NGC 4993 is projected to be this old \cite{Blanchard:2017csd}.

In Figure \ref{fig:loudonut}, the per-nucleon cross-section sensitivity obtainable for heavy, asymmetric, fermionic dark matter is shown, as calculated using the capture rate and collapse conditions presented in \cite{Bramante:2017xlb,Bramante:2015cua} and Appendix \ref{app:ADM} in this document. Lighter asymmetric dark matter can also be found using these methods, as in Refs.~\cite{Bramante:2014zca,Bramante:2015dfa,Goldman:1989nd,deLavallaz:2010wp,Kouvaris:2010jy,McDermott:2011jp,Bramante:2013hn,Bell:2013xk,
Bertoni:2013bsa,Guver:2012ba,Kouvaris:2013kra,Kurita:2015vga}.

\section{Fast radio bursts from dark matter}
\label{sec:frbs}
\begin{figure*}[t!]
\includegraphics[width=0.45\textwidth]{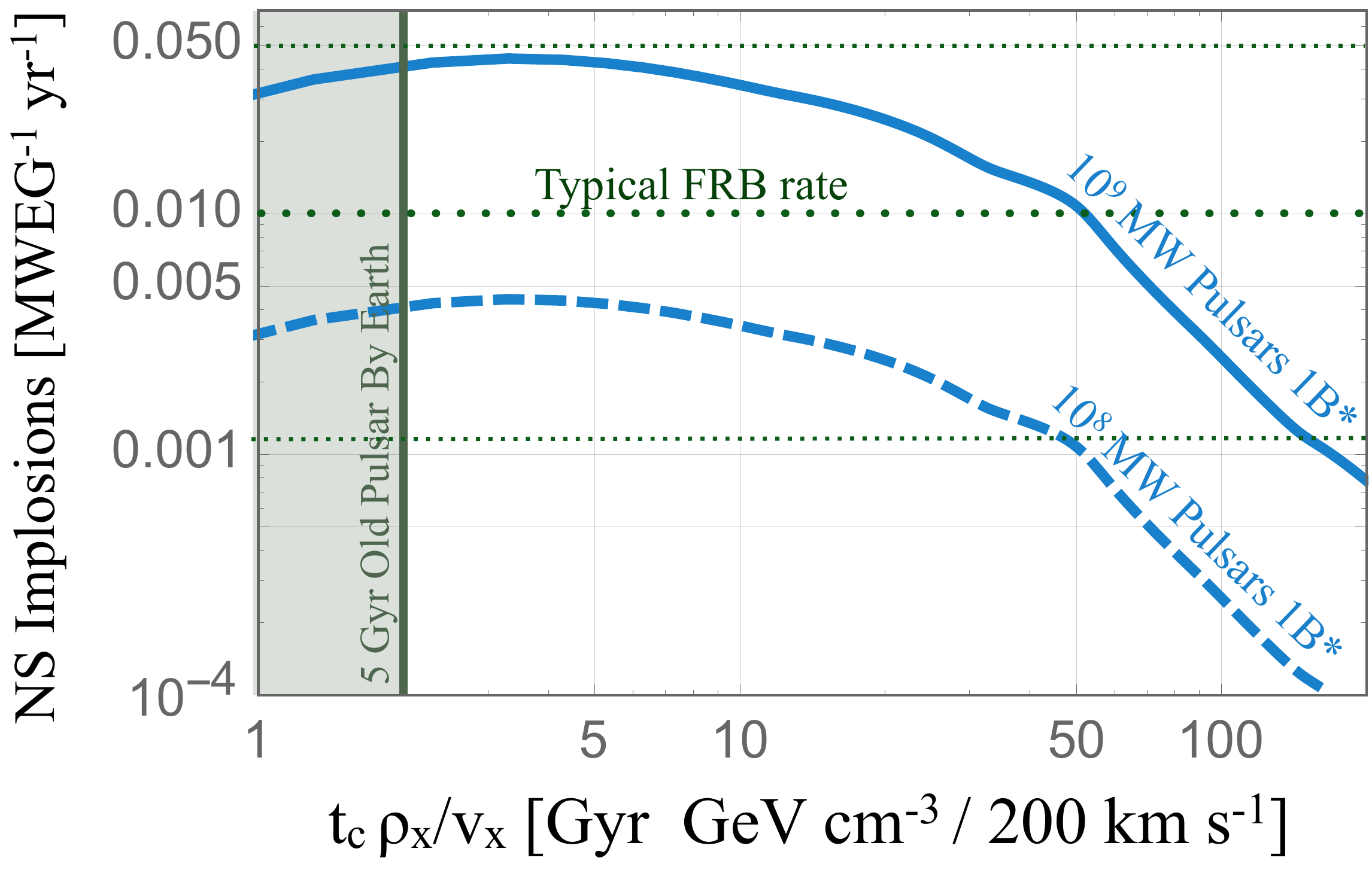}~~~
\includegraphics[width=0.45\textwidth]{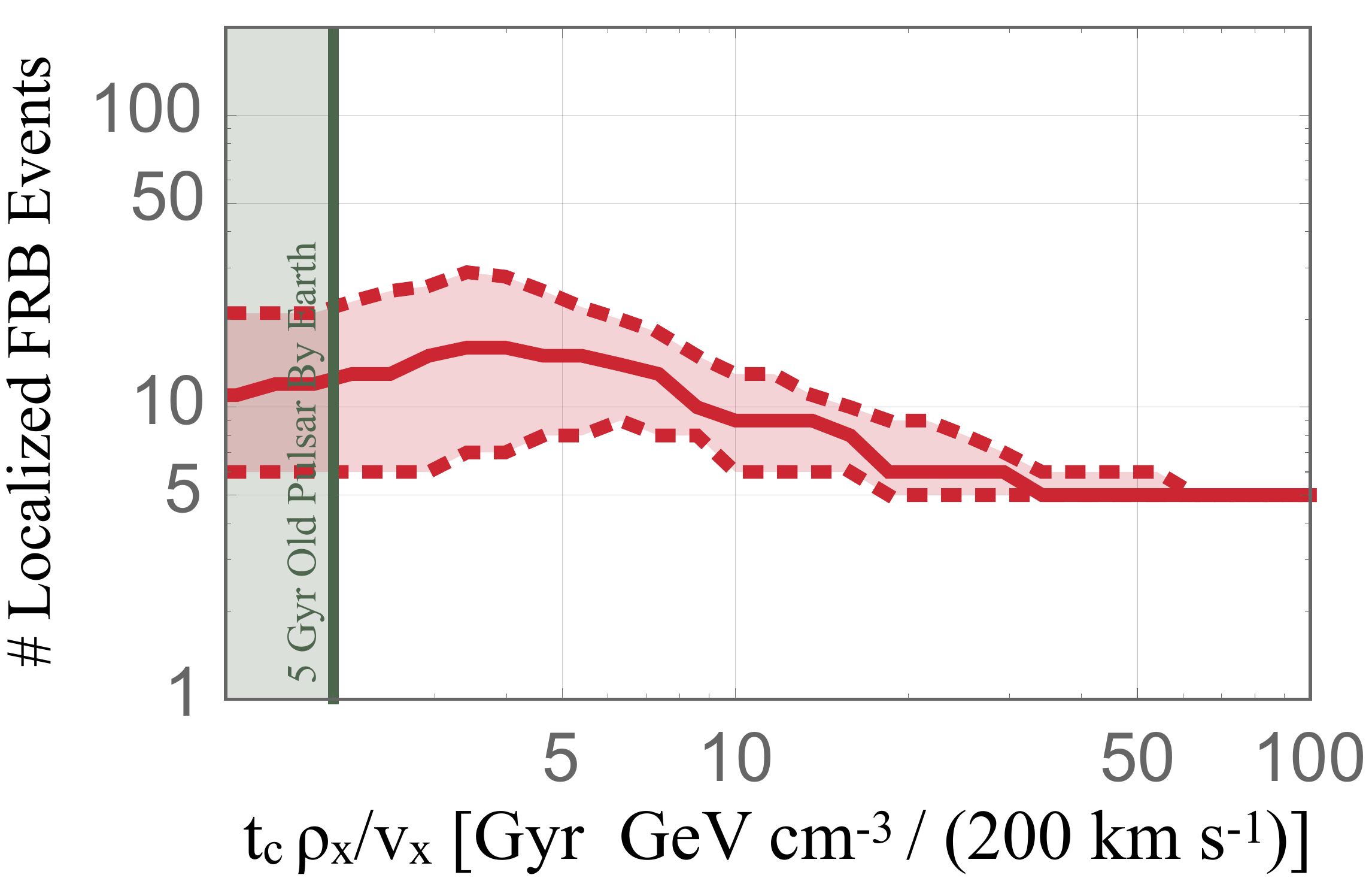}
\caption{\emph{Left}: The NS implosion rate in a Milky Way equivalent galaxy for dark matter that implodes NSs in time $t_{\rm c}$ for background dark matter density $\rho_{\rm x} $ with velocity dispersion $v_{\rm x}$ expressed in units of $\rho_{\rm x} t_{\rm c} / v_{\rm x}$. The dotted lines indicate high, median, and low fast radio burst rate estimates from surveys \cite{Rane:2015sxa,Wiel:2016pdl}. Implosion rates are shown for $10^8$ and $10^9$ NSs with a 1B* spatial distribution \cite{2010A&A...510A..23S}. \emph{Right}: Number of fast radio bursts, localized to $\sim 1 {~ \rm kpc}$ in a host galaxy, required to test whether fast radio bursts originate from NS-imploding dark matter, against the hypothesis that FRBs come from a non-imploding population of NSs, at $2 \sigma$ significance. }
\label{fig:implorate}
\end{figure*}
Fast radio bursts (FRBs) are a newly-discovered class of millisecond-length $\sim$Ghz radio pulses found to distances of 2 Gpc with an all sky rate $\sim 10^4/$day, whose source is unknown \cite{Lorimer:2007qn,Thornton:2013iua}. In \cite{Fuller:2014rza} it was proposed that dark matter-induced NS implosions may be the source of FRBs. The electromagnetic energy released by a NS implosion matches what is required for an FRB \cite{Palenzuela:2012my,Dionysopoulou:2012zv,Falcke:2013xpa}. Ref.~\cite{Fuller:2014rza} calculated the per MWEG NS implosion rate assuming a constant star formation history and a NS population that tracked the baryonic density in an MWEG. We improve on these rate calculations by using a realistic star formation history \cite{Hopkins:2006bw} and NS distribution \cite{2010A&A...510A..23S}, see Figure \ref{fig:implorate}.

The rate of NS collapse due to dark matter accumulation in the Milky Way can be estimated in several limits. In the limit of arbitrarily rapid dark matter accumulation, all NSs collapse to BHs soon after formation and the NS collapse rate is equal to the supernova rate (approximately 0.02~yr$^{-1}$~\citep{Diehl:2006cf}). This scenario is ruled out by observations of Gyr old neutron stars. Intriguingly, in the case of less rapid dark matter-induced NS implosions, the overall present-day NS implosion rate actually increases due to enhanced star formation rate in the \emph{young} Milky Way. While the star-formation rate of the Milky Way is currently $\sim$0.68 --- 1.45~M$_\odot$~yr$^{-1}$~\cite{2010ApJ...710L..11R}, the rate at redshift z=2 was approximately 14~M$_\odot$~yr$^{-1}$~\cite{vanDokkum:2013hza, 2014ApJ...781L..31S, 2015MNRAS.451.4223M}. If dark matter induces most $\sim 10$ Gyr old NSs to implode in the present epoch, then the current implosion rate will reflect the high-redshift star formation rate. Because the dark matter density diminishes with increasing distance from the galactic center, the delay-time between NS formation and NS collapse varies considerably as a function of galactocentric radius, as evident in Figure 1 in the main text. In Figure \ref{fig:implorate} we plot the number of NS implosions expected per MWEG as a function of $t_{\rm c} \rho_{\rm x} /v_{\rm x}$. Upper, median, and lower  FRB rates are also indicated for comparison \cite{Rane:2015sxa,Wiel:2016pdl}.

\section{Gravitational waves from a neutron star implosion in the Milky Way}
\label{sec:conc}
We have identified new signatures of neutron star-imploding dark matter, and fashioned qualitatively new methods for uncovering this dark sector using imminent astronomical observations. Specifically, our proposed analysis of NS merger kilonova locations has the potential to explore dark matter-nucleon scattering cross-sections up to ten orders of magnitude beyond present direct detection experiments. Finally, we note that the collapse of a neutron star into a BH could be detected directly at advanced LV, if the NS resides in the Milky Way. As we have calculated in Section 2, NS implosion event rates may be as large as $0.05$ per year. Reference \cite{Baiotti:2007np} finds the following strain for a NS collapsing to a BH,
$
h_c \sim 5 \times 10^{-22} \left( \frac{M}{M_\odot} \right) \left( \frac{\rm 10 ~kpc}{D} \right)  ~@~ 531~{\rm Hz},
$
so that advanced LV \cite{Aasi:2013wya} may find an implosion out to $\sim 1~{\rm Mpc}$. We leave additional gravitational signatures of NS-imploding dark matter to future work, along with the application of the spatial kilonova analysis introduced here, to electromagnetic transients from exotic compact object mergers \cite{Giudice:2016zpa,Bramante:2017xlb}.
\\

\acknowledgments We thank Asimina Arvanitaki, Masha Baryakhtar, Shaon Ghosh, Robert Lasenby, Shirley Weishi Li, and Luis Lehner for useful discussions and correspondence. Research at Perimeter Institute is supported by the Government of Canada through Industry Canada and by the Province of Ontario through the Ministry of Economic Development \& Innovation. J.~B.~thanks the Aspen Center for Physics, which is supported by National Science Foundation grant PHY-1066293. T.~L.~is partially supported by NSF Grant PHY-1404311 to John Beacom. 

\appendix

\section{Cumulative Distribution Functions}
\label{app:cdfs}
\begin{figure*}[t!]
\includegraphics[width=0.95\textwidth]{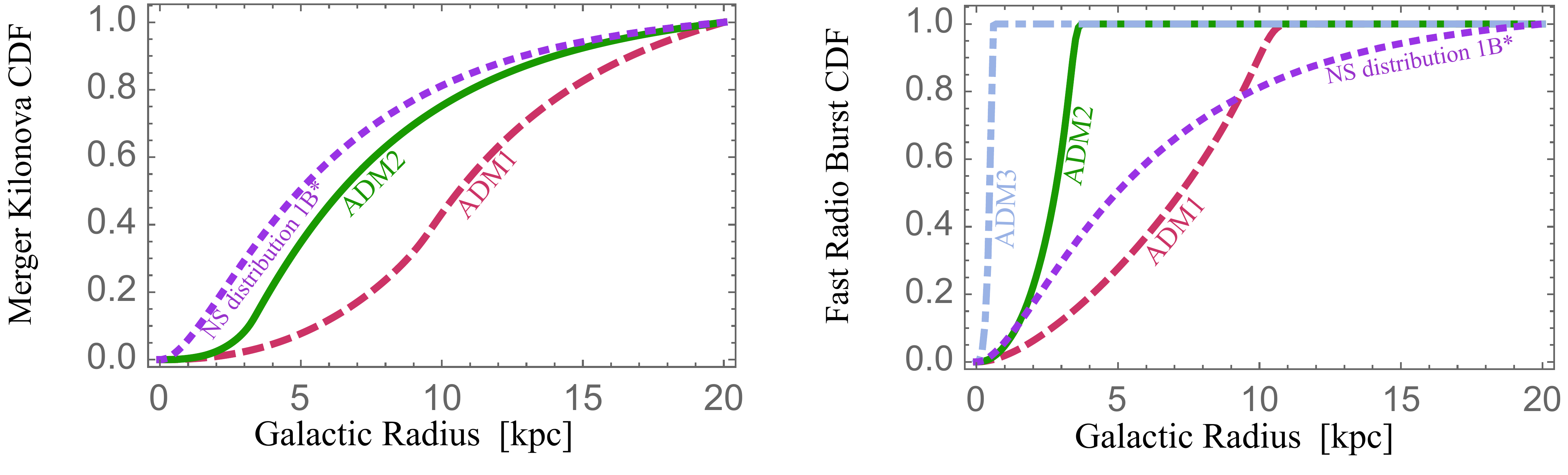}
\caption{Cumulative distribution functions corresponding to the Kolmogorov-Smirnov tests shown in  Figures \ref{fig:loudonut} and \ref{fig:implorate} are given, for merger kilonovae (left) and fast radio bursts (right). Both merger kilonovae and FRBs are assumed to follow the distribution of NSs in an MWEG (dotted purple line labeled 1B*). This distribution can be compared with distributions for representative NS-imploding dark matter models, ADM1, ADM2, and ADM3, defined by $t_{\rm c} \rho_{\rm x} /v_{\rm x}  = 3$, $15$, and $100$ $\rm Gyr~GeV/cm^3~ (200 ~km/s)^{-1}$ respectively, see Eq. (2) of the main text.}
\label{fig:cdfs}
\end{figure*}

Figure \ref{fig:cdfs} shows the cumulative distribution functions used in the Kolmogorov-Smirnov tests, whose results are displayed in Figures \ref{fig:loudonut} and \ref{fig:implorate}. As indicated, the displayed CDFs here were generated using the neutron star distribution model 1B* in \cite{2010A&A...510A..23S}. However, note that we have found that Kolmogorov-Smirnov tests utilizing other pulsar distribution models in \cite{2010A&A...510A..23S} require a similar number of NS mergers to achieve sensitivity to asymmetric dark matter comparable to the results shown in  Figures \ref{fig:loudonut} and \ref{fig:implorate}.

\section{Dark matter-induced Neutron Star Implosion Time}
\label{app:ADM}

The analysis of NS implosions in Section 1 assumed that the longest timescale in the NS implosion process is the time for a NS to accrete a BH-forming mass of dark matter. Here we justify this assumption by computing timescales for all dynamical processes leading up to dark matter-induced NS implosions. As an example, we consider fermionic dark matter in the PeV mass range with dark matter-nucleon cross-section $\sigma_{n\rm x} \sim 10^{-45}~ {\rm cm^2}$. For $\sigma_{n\rm x} \gtrsim 10^{-45}~ {\rm cm^2}$, all PeV mass dark matter passing through the NS will be captured \cite{Bramante:2017xlb}, as assumed in Section 1. Extending this work to PeV mass bosonic dark matter requires minor modifications discussed around Eq. (3) of the main text, with similar results obtained. Similar results arises for low mass dark matter \cite{Kouvaris:2011gb, deLavallaz:2010wp, Bramante:2013hn, Bertoni:2013bsa, Bramante:2015cua}, where sensitivity to $\sigma_{\rm nx}$ scales inversely with $t_{\rm c} \rho_{\rm x} /v_{\rm x} $ -- so that the $t_{\rm c} \rho_{\rm x} /v_{\rm x} $ reach of this study results in up to a $100$-fold improvement in $\sigma_{\rm nx}$ sensitivity.

A small BH may form when a NS accumulates so much dark matter, that the dark matter cannot support its own weight with degeneracy pressure. This critical mass for fermionic dark matter is $M_{\rm crit}^{\rm f} \sim \frac{m_{\rm pl}^3}{m_{\rm x}^2}$, as reported in the main text. In principle ($e.g.$ for lower mass dark matter) dark matter may accumulate to $M_{\rm crit}^{\rm f}$ size in a NS, yet not implode. This is because until dark matter in a NS ``self-gravitates" or equivalently forms a bulk whose density exceeds the NS density, it will remain stable.\footnote{In the case of dark matter with substantial self-interactions, this computation is different \cite{Kouvaris:2011gb,Bramante:2013nma,Bramante:2015dfa}.} We find the self-gravitating mass $M_{\rm sg}$ \cite{Bramante:2013hn} for PeV mass dark matter, and determine that $M_{\rm crit} \gg M_{\rm sg}$, which justifies our assumption in the main text, that a BH will form once $M_{\rm crit}$ dark matter accumulates. For the limiting case of a younger NS with temperature $T_{\rm NS} \sim 10^5~{\rm K}$, 
\begin{align}
M_{sg} \simeq 5 \times 10^{37} \:{\rm GeV} \left(\frac{T_{\rm NS}}{10^5~ {\rm K}}\right)^{3/2}
\left(\frac{\rm PeV}{m_X}\right)^{3/2},
\end{align}
and one can see that $M_{\rm crit}$ is at least $10^6$ times larger than $M_{sg}$ for PeV mass dark matter.

Next we review the dynamical timescales for a NS to become converted to a BH by accumulated dark matter. First the dark matter particles thermally equilibrate with the neutron star, through repeated scattering; we denote this thermalization time scale by $\tau_{\rm th}$. Once an unstable mass $M_{\rm crit}$ of dark matter has thermalized into a small volume, the dark matter will collapse, cool, and form a BH, over a time $\tau_{\rm co}$. Lastly, the small BH formed of dark matter accretes the surrounding neutron star in a time $\tau_{Bondi}$. We will see that each process occurs much faster than the $t_c \gtrsim {\rm Myr}$ time scale, and conclude that the neutron star implosion time is determined by $t_c$.

We first consider the thermalization. Dark matter particles captured and accumulated in the neutron star thermalize with the neutrons and cool to temperature $T\simeq 10^5 \rm\:K$, same as the neutron star, before it start lose more kinetic energy and eventually collapse into a BH. The time scale $\tau_{th}$ is determined by the neutron-dark matter collisions \cite{Bertoni:2013bsa}, in particular
\begin{align}
\tau_{\rm th} 
\simeq
8 \times 10^{-3} {\:\rm yr}\left(\frac{m_{\rm x}}{{\rm PeV}}\right)
\left(\frac{10^{-45}{\rm\:cm^2}}{\sigma_{nx}}\right)\left(\frac{10^5 {\rm\:K}}{T_{\rm NS}}\right)^2.
\end{align}

After thermalization, the dark matter particles form a spherical configuration of radius $r_0\sim (9T/8\pi G \rho_c m_{\rm x})^{1/2}$, where the dark matter density at collapse is equal to the NS density $\rho_c \sim \rho_{\rm n}$.
The dark matter particles can further lose more of their energy and collapse into a BH. There are several mechanisms that contribute to this cooling, with associated time scales. Here we focus on cooling via dark matter-neutron scattering, while other cooling mechanisms can be found in {\em e.g.} \cite{Goldman:1989nd}. The dark matter cooling and collapse time is approximately the time for dark matter to lose $\mathcal{O}(1)$ of its kinetic energy to surrounding neutrons,
\begin{align}
\tau_{\rm co} & \simeq \frac{1}{n \sigma_{n \rm x}v_{\rm xc}} \left(\frac{p_F}{\Delta p}\right) \left(\frac{m_{\rm x}}{2 m_n}\right)\label{eq:tau_co}
\nonumber \\
&\simeq 4 \times 10^5~ {\rm yrs} \left(\frac{m_{\rm x}}{{\rm PeV}}\right)\left(\frac{10^{-45}~\rm cm^2}{\sigma_{\rm n  x}}\right) \left(\frac{r_x}{r_0}\right),
\end{align}
where $n$ is the number density of the neutrons.
The first term $1/n \sigma_{n \rm x} v_{\rm xc}$ is the time for a single collision. In addition, Pauli blocking has to be taken into account, as it reduces cross-section by a factor of $\Delta p /p_F$, hence the second term. Here $p_F\sim 0.5~ {\rm GeV}$ is the neutron Fermi momentum in a NS and $\Delta p\sim m_n v_{\rm xc}$. The factor of $v_{\rm xc}$ here is the velocity of the dark matter sphere as it collapses through radius $r_x$, which can be written as $v_{\rm xc}\sim (G m_{\rm x}/r_{x})^{1/2}$. 
The final factor takes into account that in each collision only a fraction $\sim 2 m_n / m_{\rm x}$ of the dark matter kinetic energy is transferred, so one requires $\sim\left(\frac{m_{\rm x}}{2 m_n}\right)$ collisions for an order-one energy loss. 

With the BH formed, assuming it accretes the remainder of the NS, the time for which depends on the BHBH mass $M_{\rm crit}$ ($e.g.$ \cite{Bramante:2016mzo,Kouvaris:2013kra}). 
\begin{align}
\tau_{\rm Bondi} \sim 0.1 {\rm\:yrs} \left(\frac{m_X}{\rm PeV}\right)^2 
\end{align}
We find that $\tau_{\rm th},\tau_{\rm co}, \tau_{\rm Bondi}$ are much shorter than $\sim t_c$, which therefore determines the time until a NS implodes.

\bibliography{dmerger}

\begin{thebibliography}{71}%
\makeatletter
\providecommand \@ifxundefined [1]{%
 \@ifx{#1\undefined}
}%
\providecommand \@ifnum [1]{%
 \ifnum #1\expandafter \@firstoftwo
 \else \expandafter \@secondoftwo
 \fi
}%
\providecommand \@ifx [1]{%
 \ifx #1\expandafter \@firstoftwo
 \else \expandafter \@secondoftwo
 \fi
}%
\providecommand \natexlab [1]{#1}%
\providecommand \enquote  [1]{``#1''}%
\providecommand \bibnamefont  [1]{#1}%
\providecommand \bibfnamefont [1]{#1}%
\providecommand \citenamefont [1]{#1}%
\providecommand \href@noop [0]{\@secondoftwo}%
\providecommand \href [0]{\begingroup \@sanitize@url \@href}%
\providecommand \@href[1]{\@@startlink{#1}\@@href}%
\providecommand \@@href[1]{\endgroup#1\@@endlink}%
\providecommand \@sanitize@url [0]{\catcode `\\12\catcode `\$12\catcode
  `\&12\catcode `\#12\catcode `\^12\catcode `\_12\catcode `\%12\relax}%
\providecommand \@@startlink[1]{}%
\providecommand \@@endlink[0]{}%
\providecommand \url  [0]{\begingroup\@sanitize@url \@url }%
\providecommand \@url [1]{\endgroup\@href {#1}{\urlprefix }}%
\providecommand \urlprefix  [0]{URL }%
\providecommand \Eprint [0]{\href }%
\providecommand \doibase [0]{http://dx.doi.org/}%
\providecommand \selectlanguage [0]{\@gobble}%
\providecommand \bibinfo  [0]{\@secondoftwo}%
\providecommand \bibfield  [0]{\@secondoftwo}%
\providecommand \translation [1]{[#1]}%
\providecommand \BibitemOpen [0]{}%
\providecommand \bibitemStop [0]{}%
\providecommand \bibitemNoStop [0]{.\EOS\space}%
\providecommand \EOS [0]{\spacefactor3000\relax}%
\providecommand \BibitemShut  [1]{\csname bibitem#1\endcsname}%
\let\auto@bib@innerbib\@empty
\bibitem [{\citenamefont {Bramante}\ and\ \citenamefont
  {Linden}(2014)}]{Bramante:2014zca}%
  \BibitemOpen
  \bibfield  {author} {\bibinfo {author} {\bibfnamefont {Joseph}\ \bibnamefont
  {Bramante}}\ and\ \bibinfo {author} {\bibfnamefont {Tim}\ \bibnamefont
  {Linden}},\ }\bibfield  {title} {\enquote {\bibinfo {title} {{Detecting Dark
  Matter with Imploding Pulsars in the Galactic Center}},}\ }\href {\doibase
  10.1103/PhysRevLett.113.191301} {\bibfield  {journal} {\bibinfo  {journal}
  {Phys. Rev. Lett.}\ }\textbf {\bibinfo {volume} {113}},\ \bibinfo {pages}
  {191301} (\bibinfo {year} {2014})},\ \Eprint {http://arxiv.org/abs/1405.1031}
  {arXiv:1405.1031 [astro-ph.HE]} \BibitemShut {NoStop}%
\bibitem [{\citenamefont {Bramante}\ and\ \citenamefont
  {Elahi}(2015)}]{Bramante:2015dfa}%
  \BibitemOpen
  \bibfield  {author} {\bibinfo {author} {\bibfnamefont {Joseph}\ \bibnamefont
  {Bramante}}\ and\ \bibinfo {author} {\bibfnamefont {Fatemeh}\ \bibnamefont
  {Elahi}},\ }\bibfield  {title} {\enquote {\bibinfo {title} {{Higgs portals to
  pulsar collapse}},}\ }\href {\doibase 10.1103/PhysRevD.91.115001} {\bibfield
  {journal} {\bibinfo  {journal} {Phys. Rev.}\ }\textbf {\bibinfo {volume}
  {D91}},\ \bibinfo {pages} {115001} (\bibinfo {year} {2015})},\ \Eprint
  {http://arxiv.org/abs/1504.04019} {arXiv:1504.04019 [hep-ph]} \BibitemShut
  {NoStop}%
\bibitem [{\citenamefont {Goldman}\ and\ \citenamefont
  {Nussinov}(1989)}]{Goldman:1989nd}%
  \BibitemOpen
  \bibfield  {author} {\bibinfo {author} {\bibfnamefont {I.}~\bibnamefont
  {Goldman}}\ and\ \bibinfo {author} {\bibfnamefont {S.}~\bibnamefont
  {Nussinov}},\ }\bibfield  {title} {\enquote {\bibinfo {title} {{Weakly
  Interacting Massive Particles and Neutron Stars}},}\ }\href {\doibase
  10.1103/PhysRevD.40.3221} {\bibfield  {journal} {\bibinfo  {journal} {Phys.
  Rev.}\ }\textbf {\bibinfo {volume} {D40}},\ \bibinfo {pages} {3221--3230}
  (\bibinfo {year} {1989})}\BibitemShut {NoStop}%
\bibitem [{\citenamefont {de~Lavallaz}\ and\ \citenamefont
  {Fairbairn}(2010)}]{deLavallaz:2010wp}%
  \BibitemOpen
  \bibfield  {author} {\bibinfo {author} {\bibfnamefont {Arnaud}\ \bibnamefont
  {de~Lavallaz}}\ and\ \bibinfo {author} {\bibfnamefont {Malcolm}\ \bibnamefont
  {Fairbairn}},\ }\bibfield  {title} {\enquote {\bibinfo {title} {{Neutron
  Stars as Dark Matter Probes}},}\ }\href {\doibase 10.1103/PhysRevD.81.123521}
  {\bibfield  {journal} {\bibinfo  {journal} {Phys. Rev.}\ }\textbf {\bibinfo
  {volume} {D81}},\ \bibinfo {pages} {123521} (\bibinfo {year} {2010})},\
  \Eprint {http://arxiv.org/abs/1004.0629} {arXiv:1004.0629 [astro-ph.GA]}
  \BibitemShut {NoStop}%
\bibitem [{\citenamefont {Kouvaris}\ and\ \citenamefont
  {Tinyakov}(2011)}]{Kouvaris:2010jy}%
  \BibitemOpen
  \bibfield  {author} {\bibinfo {author} {\bibfnamefont {Chris}\ \bibnamefont
  {Kouvaris}}\ and\ \bibinfo {author} {\bibfnamefont {Peter}\ \bibnamefont
  {Tinyakov}},\ }\bibfield  {title} {\enquote {\bibinfo {title} {{Constraining
  Asymmetric Dark Matter through observations of compact stars}},}\ }\href
  {\doibase 10.1103/PhysRevD.83.083512} {\bibfield  {journal} {\bibinfo
  {journal} {Phys. Rev.}\ }\textbf {\bibinfo {volume} {D83}},\ \bibinfo {pages}
  {083512} (\bibinfo {year} {2011})},\ \Eprint {http://arxiv.org/abs/1012.2039}
  {arXiv:1012.2039 [astro-ph.HE]} \BibitemShut {NoStop}%
\bibitem [{\citenamefont {Bramante}\ and\ \citenamefont
  {Linden}(2016)}]{Bramante:2016mzo}%
  \BibitemOpen
  \bibfield  {author} {\bibinfo {author} {\bibfnamefont {Joseph}\ \bibnamefont
  {Bramante}}\ and\ \bibinfo {author} {\bibfnamefont {Tim}\ \bibnamefont
  {Linden}},\ }\bibfield  {title} {\enquote {\bibinfo {title} {{On the
  $r$-Process Enrichment of Dwarf Spheroidal Galaxies}},}\ }\href {\doibase
  10.3847/0004-637X/826/1/57} {\bibfield  {journal} {\bibinfo  {journal}
  {Astrophys. J.}\ }\textbf {\bibinfo {volume} {826}},\ \bibinfo {pages} {57}
  (\bibinfo {year} {2016})},\ \Eprint {http://arxiv.org/abs/1601.06784}
  {arXiv:1601.06784 [astro-ph.HE]} \BibitemShut {NoStop}%
\bibitem [{\citenamefont {Fuller}\ and\ \citenamefont
  {Ott}(2015)}]{Fuller:2014rza}%
  \BibitemOpen
  \bibfield  {author} {\bibinfo {author} {\bibfnamefont {Jim}\ \bibnamefont
  {Fuller}}\ and\ \bibinfo {author} {\bibfnamefont {Christian}\ \bibnamefont
  {Ott}},\ }\bibfield  {title} {\enquote {\bibinfo {title} {{Dark
  Matter-induced Collapse of Neutron Stars: A Possible Link Between Fast Radio
  Bursts and the Missing Pulsar Problem}},}\ }\href {\doibase
  10.1093/mnrasl/slv049} {\bibfield  {journal} {\bibinfo  {journal} {Mon. Not.
  Roy. Astron. Soc.}\ }\textbf {\bibinfo {volume} {450}},\ \bibinfo {pages}
  {L71--L75} (\bibinfo {year} {2015})},\ \Eprint
  {http://arxiv.org/abs/1412.6119} {arXiv:1412.6119 [astro-ph.HE]} \BibitemShut
  {NoStop}%
\bibitem [{\citenamefont {Abbott}\ \emph {et~al.}(2016)\citenamefont {Abbott}
  \emph {et~al.}}]{Abbott:2016ymx}%
  \BibitemOpen
  \bibfield  {author} {\bibinfo {author} {\bibfnamefont {Benjamin~P.}\
  \bibnamefont {Abbott}} \emph {et~al.} (\bibinfo {collaboration} {Virgo, LIGO
  Scientific}),\ }\bibfield  {title} {\enquote {\bibinfo {title} {{Upper Limits
  on the Rates of Binary Neutron Star and Neutron Star - Black Hole Mergers
  From Advanced Ligo's First Observing Run}},}\ }\href {\doibase
  10.3847/2041-8205/832/2/L21} {\bibfield  {journal} {\bibinfo  {journal}
  {Astrophys. J.}\ }\textbf {\bibinfo {volume} {832}},\ \bibinfo {pages} {L21}
  (\bibinfo {year} {2016})},\ \Eprint {http://arxiv.org/abs/1607.07456}
  {arXiv:1607.07456 [astro-ph.HE]} \BibitemShut {NoStop}%
\bibitem [{\citenamefont {Doctor}\ \emph {et~al.}(2017)\citenamefont {Doctor}
  \emph {et~al.}}]{Doctor:2016gdi}%
  \BibitemOpen
  \bibfield  {author} {\bibinfo {author} {\bibfnamefont {Z.}~\bibnamefont
  {Doctor}} \emph {et~al.} (\bibinfo {collaboration} {DES}),\ }\bibfield
  {title} {\enquote {\bibinfo {title} {{A Search for Kilonovae in the Dark
  Energy Survey}},}\ }\href {\doibase 10.3847/1538-4357/aa5d09} {\bibfield
  {journal} {\bibinfo  {journal} {Astrophys. J.}\ }\textbf {\bibinfo {volume}
  {837}},\ \bibinfo {pages} {57} (\bibinfo {year} {2017})},\ \Eprint
  {http://arxiv.org/abs/1611.08052} {arXiv:1611.08052 [astro-ph.HE]}
  \BibitemShut {NoStop}%
\bibitem [{\citenamefont {Soares-Santos}\ \emph {et~al.}(2016)\citenamefont
  {Soares-Santos} \emph {et~al.}}]{Soares-Santos:2016qeb}%
  \BibitemOpen
  \bibfield  {author} {\bibinfo {author} {\bibfnamefont {M.}~\bibnamefont
  {Soares-Santos}} \emph {et~al.} (\bibinfo {collaboration} {DES}),\ }\bibfield
   {title} {\enquote {\bibinfo {title} {{A Dark Energy Camera Search for an
  Optical Counterpart to the First Advanced LIGO Gravitational Wave Event
  GW150914}},}\ }\href {\doibase 10.3847/2041-8205/823/2/L33} {\bibfield
  {journal} {\bibinfo  {journal} {Astrophys. J.}\ }\textbf {\bibinfo {volume}
  {823}},\ \bibinfo {pages} {L33} (\bibinfo {year} {2016})},\ \Eprint
  {http://arxiv.org/abs/1602.04198} {arXiv:1602.04198 [astro-ph.CO]}
  \BibitemShut {NoStop}%
\bibitem [{\citenamefont {Ghosh}\ and\ \citenamefont
  {Nelemans}(2015)}]{Ghosh:2014yga}%
  \BibitemOpen
  \bibfield  {author} {\bibinfo {author} {\bibfnamefont {Shaon}\ \bibnamefont
  {Ghosh}}\ and\ \bibinfo {author} {\bibfnamefont {Gijs}\ \bibnamefont
  {Nelemans}},\ }\bibfield  {title} {\enquote {\bibinfo {title} {{Localizing
  gravitational wave sources with optical telescopes and combining
  electromagnetic and gravitational wave data}},}\ }\href {\doibase
  10.1007/978-3-319-10488-1_5} {\bibfield  {journal} {\bibinfo  {journal}
  {Astrophys. Space Sci. Proc.}\ }\textbf {\bibinfo {volume} {40}},\ \bibinfo
  {pages} {51--58} (\bibinfo {year} {2015})},\ \Eprint
  {http://arxiv.org/abs/1406.0343} {arXiv:1406.0343 [gr-qc]} \BibitemShut
  {NoStop}%
\bibitem [{\citenamefont {Berger}\ \emph {et~al.}(2016)\citenamefont {Berger}
  \emph {et~al.}}]{Berger:2016ejd}%
  \BibitemOpen
  \bibfield  {author} {\bibinfo {author} {\bibfnamefont {Philippe}\
  \bibnamefont {Berger}} \emph {et~al.},\ }\bibfield  {title} {\enquote
  {\bibinfo {title} {{Holographic Beam Mapping of the CHIME Pathfinder
  Array}},}\ }\href {\doibase 10.1117/12.2233782} {\bibfield  {journal}
  {\bibinfo  {journal} {Proc. SPIE Int. Soc. Opt. Eng.}\ }\textbf {\bibinfo
  {volume} {9906}},\ \bibinfo {pages} {99060D} (\bibinfo {year} {2016})},\
  \Eprint {http://arxiv.org/abs/1607.01473} {arXiv:1607.01473 [astro-ph.IM]}
  \BibitemShut {NoStop}%
\bibitem [{\citenamefont {Newburgh}\ \emph {et~al.}(2016)\citenamefont
  {Newburgh} \emph {et~al.}}]{Newburgh:2016mwi}%
  \BibitemOpen
  \bibfield  {author} {\bibinfo {author} {\bibfnamefont {L.~B.}\ \bibnamefont
  {Newburgh}} \emph {et~al.},\ }\bibfield  {title} {\enquote {\bibinfo {title}
  {{HIRAX: A Probe of Dark Energy and Radio Transients}},}\ }\href {\doibase
  10.1117/12.2234286} {\bibfield  {journal} {\bibinfo  {journal} {Proc. SPIE
  Int. Soc. Opt. Eng.}\ }\textbf {\bibinfo {volume} {9906}},\ \bibinfo {pages}
  {99065X} (\bibinfo {year} {2016})},\ \Eprint
  {http://arxiv.org/abs/1607.02059} {arXiv:1607.02059 [astro-ph.IM]}
  \BibitemShut {NoStop}%
\bibitem [{\citenamefont {Abbott}\ \emph
  {et~al.}(2017{\natexlab{a}})\citenamefont {Abbott} \emph
  {et~al.}}]{TheLIGOScientific:2017qsa}%
  \BibitemOpen
  \bibfield  {author} {\bibinfo {author} {\bibfnamefont {Benjamin~P.}\
  \bibnamefont {Abbott}} \emph {et~al.} (\bibinfo {collaboration} {Virgo, LIGO
  Scientific}),\ }\bibfield  {title} {\enquote {\bibinfo {title} {{GW170817:
  Observation of Gravitational Waves from a Binary Neutron Star Inspiral}},}\
  }\href {\doibase 10.1103/PhysRevLett.119.161101} {\bibfield  {journal}
  {\bibinfo  {journal} {Phys. Rev. Lett.}\ }\textbf {\bibinfo {volume} {119}},\
  \bibinfo {pages} {161101} (\bibinfo {year} {2017}{\natexlab{a}})},\ \Eprint
  {http://arxiv.org/abs/1710.05832} {arXiv:1710.05832 [gr-qc]} \BibitemShut
  {NoStop}%
\bibitem [{\citenamefont {Abbott}\ \emph
  {et~al.}(2017{\natexlab{b}})\citenamefont {Abbott} \emph
  {et~al.}}]{GBM:2017lvd}%
  \BibitemOpen
  \bibfield  {author} {\bibinfo {author} {\bibfnamefont {B.~P.}\ \bibnamefont
  {Abbott}} \emph {et~al.} (\bibinfo {collaboration} {GROND, SALT Group,
  OzGrav, DFN, INTEGRAL, Virgo, Insight-Hxmt, MAXI Team, Fermi-LAT, J-GEM,
  RATIR, IceCube, CAASTRO, LWA, ePESSTO, GRAWITA, RIMAS, SKA South
  Africa/MeerKAT, H.E.S.S., 1M2H Team, IKI-GW Follow-up, Fermi GBM, Pi of Sky,
  DWF (Deeper Wider Faster Program), Dark Energy Survey, MASTER, AstroSat
  Cadmium Zinc Telluride Imager Team, Swift, Pierre Auger, ASKAP, VINROUGE,
  JAGWAR, Chandra Team at McGill University, TTU-NRAO, GROWTH, AGILE Team, MWA,
  ATCA, AST3, TOROS, Pan-STARRS, NuSTAR, ATLAS Telescopes, BOOTES, CaltechNRAO,
  LIGO Scientific, High Time Resolution Universe Survey, Nordic Optical
  Telescope, Las Cumbres Observatory Group, TZAC Consortium, LOFAR, IPN, DLT40,
  Texas Tech University, HAWC, ANTARES, KU, Dark Energy Camera GW-EM, CALET,
  Euro VLBI Team, ALMA}),\ }\bibfield  {title} {\enquote {\bibinfo {title}
  {{Multi-messenger Observations of a Binary Neutron Star Merger}},}\ }\href
  {\doibase 10.3847/2041-8213/aa91c9} {\bibfield  {journal} {\bibinfo
  {journal} {Astrophys. J.}\ }\textbf {\bibinfo {volume} {848}},\ \bibinfo
  {pages} {L12} (\bibinfo {year} {2017}{\natexlab{b}})},\ \Eprint
  {http://arxiv.org/abs/1710.05833} {arXiv:1710.05833 [astro-ph.HE]}
  \BibitemShut {NoStop}%
\bibitem [{\citenamefont {Petraki}\ and\ \citenamefont
  {Volkas}(2013)}]{Petraki:2013wwa}%
  \BibitemOpen
  \bibfield  {author} {\bibinfo {author} {\bibfnamefont {Kalliopi}\
  \bibnamefont {Petraki}}\ and\ \bibinfo {author} {\bibfnamefont {Raymond~R.}\
  \bibnamefont {Volkas}},\ }\bibfield  {title} {\enquote {\bibinfo {title}
  {{Review of asymmetric dark matter}},}\ }\href {\doibase
  10.1142/S0217751X13300287} {\bibfield  {journal} {\bibinfo  {journal} {Int.
  J. Mod. Phys.}\ }\textbf {\bibinfo {volume} {A28}},\ \bibinfo {pages}
  {1330028} (\bibinfo {year} {2013})},\ \Eprint
  {http://arxiv.org/abs/1305.4939} {arXiv:1305.4939 [hep-ph]} \BibitemShut
  {NoStop}%
\bibitem [{\citenamefont {Zurek}(2014)}]{Zurek:2013wia}%
  \BibitemOpen
  \bibfield  {author} {\bibinfo {author} {\bibfnamefont {Kathryn~M.}\
  \bibnamefont {Zurek}},\ }\bibfield  {title} {\enquote {\bibinfo {title}
  {{Asymmetric Dark Matter: Theories, Signatures, and Constraints}},}\ }\href
  {\doibase 10.1016/j.physrep.2013.12.001} {\bibfield  {journal} {\bibinfo
  {journal} {Phys. Rept.}\ }\textbf {\bibinfo {volume} {537}},\ \bibinfo
  {pages} {91--121} (\bibinfo {year} {2014})},\ \Eprint
  {http://arxiv.org/abs/1308.0338} {arXiv:1308.0338 [hep-ph]} \BibitemShut
  {NoStop}%
\bibitem [{\citenamefont {McDermott}\ \emph {et~al.}(2012)\citenamefont
  {McDermott}, \citenamefont {Yu},\ and\ \citenamefont
  {Zurek}}]{McDermott:2011jp}%
  \BibitemOpen
  \bibfield  {author} {\bibinfo {author} {\bibfnamefont {Samuel~D.}\
  \bibnamefont {McDermott}}, \bibinfo {author} {\bibfnamefont {Hai-Bo}\
  \bibnamefont {Yu}}, \ and\ \bibinfo {author} {\bibfnamefont {Kathryn~M.}\
  \bibnamefont {Zurek}},\ }\bibfield  {title} {\enquote {\bibinfo {title}
  {{Constraints on Scalar Asymmetric Dark Matter from Black Hole Formation in
  Neutron Stars}},}\ }\href {\doibase 10.1103/PhysRevD.85.023519} {\bibfield
  {journal} {\bibinfo  {journal} {Phys. Rev.}\ }\textbf {\bibinfo {volume}
  {D85}},\ \bibinfo {pages} {023519} (\bibinfo {year} {2012})},\ \Eprint
  {http://arxiv.org/abs/1103.5472} {arXiv:1103.5472 [hep-ph]} \BibitemShut
  {NoStop}%
\bibitem [{\citenamefont {Bramante}\ \emph {et~al.}(2013)\citenamefont
  {Bramante}, \citenamefont {Fukushima},\ and\ \citenamefont
  {Kumar}}]{Bramante:2013hn}%
  \BibitemOpen
  \bibfield  {author} {\bibinfo {author} {\bibfnamefont {Joseph}\ \bibnamefont
  {Bramante}}, \bibinfo {author} {\bibfnamefont {Keita}\ \bibnamefont
  {Fukushima}}, \ and\ \bibinfo {author} {\bibfnamefont {Jason}\ \bibnamefont
  {Kumar}},\ }\bibfield  {title} {\enquote {\bibinfo {title} {{Constraints on
  bosonic dark matter from observation of old neutron stars}},}\ }\href
  {\doibase 10.1103/PhysRevD.87.055012} {\bibfield  {journal} {\bibinfo
  {journal} {Phys. Rev.}\ }\textbf {\bibinfo {volume} {D87}},\ \bibinfo {pages}
  {055012} (\bibinfo {year} {2013})},\ \Eprint {http://arxiv.org/abs/1301.0036}
  {arXiv:1301.0036 [hep-ph]} \BibitemShut {NoStop}%
\bibitem [{\citenamefont {Bell}\ \emph {et~al.}(2013)\citenamefont {Bell},
  \citenamefont {Melatos},\ and\ \citenamefont {Petraki}}]{Bell:2013xk}%
  \BibitemOpen
  \bibfield  {author} {\bibinfo {author} {\bibfnamefont {Nicole~F.}\
  \bibnamefont {Bell}}, \bibinfo {author} {\bibfnamefont {Andrew}\ \bibnamefont
  {Melatos}}, \ and\ \bibinfo {author} {\bibfnamefont {Kalliopi}\ \bibnamefont
  {Petraki}},\ }\bibfield  {title} {\enquote {\bibinfo {title} {{Realistic
  neutron star constraints on bosonic asymmetric dark matter}},}\ }\href
  {\doibase 10.1103/PhysRevD.87.123507} {\bibfield  {journal} {\bibinfo
  {journal} {Phys. Rev.}\ }\textbf {\bibinfo {volume} {D87}},\ \bibinfo {pages}
  {123507} (\bibinfo {year} {2013})},\ \Eprint {http://arxiv.org/abs/1301.6811}
  {arXiv:1301.6811 [hep-ph]} \BibitemShut {NoStop}%
\bibitem [{\citenamefont {Bertoni}\ \emph {et~al.}(2013)\citenamefont
  {Bertoni}, \citenamefont {Nelson},\ and\ \citenamefont
  {Reddy}}]{Bertoni:2013bsa}%
  \BibitemOpen
  \bibfield  {author} {\bibinfo {author} {\bibfnamefont {Bridget}\ \bibnamefont
  {Bertoni}}, \bibinfo {author} {\bibfnamefont {Ann~E.}\ \bibnamefont
  {Nelson}}, \ and\ \bibinfo {author} {\bibfnamefont {Sanjay}\ \bibnamefont
  {Reddy}},\ }\bibfield  {title} {\enquote {\bibinfo {title} {{Dark Matter
  Thermalization in Neutron Stars}},}\ }\href {\doibase
  10.1103/PhysRevD.88.123505} {\bibfield  {journal} {\bibinfo  {journal} {Phys.
  Rev.}\ }\textbf {\bibinfo {volume} {D88}},\ \bibinfo {pages} {123505}
  (\bibinfo {year} {2013})},\ \Eprint {http://arxiv.org/abs/1309.1721}
  {arXiv:1309.1721 [hep-ph]} \BibitemShut {NoStop}%
\bibitem [{\citenamefont {Güver}\ \emph {et~al.}(2014)\citenamefont {Güver},
  \citenamefont {Erkoca}, \citenamefont {Hall~Reno},\ and\ \citenamefont
  {Sarcevic}}]{Guver:2012ba}%
  \BibitemOpen
  \bibfield  {author} {\bibinfo {author} {\bibfnamefont {Tolga}\ \bibnamefont
  {Güver}}, \bibinfo {author} {\bibfnamefont {Arif~Emre}\ \bibnamefont
  {Erkoca}}, \bibinfo {author} {\bibfnamefont {Mary}\ \bibnamefont
  {Hall~Reno}}, \ and\ \bibinfo {author} {\bibfnamefont {Ina}\ \bibnamefont
  {Sarcevic}},\ }\bibfield  {title} {\enquote {\bibinfo {title} {{On the
  capture of dark matter by neutron stars}},}\ }\href {\doibase
  10.1088/1475-7516/2014/05/013} {\bibfield  {journal} {\bibinfo  {journal}
  {JCAP}\ }\textbf {\bibinfo {volume} {1405}},\ \bibinfo {pages} {013}
  (\bibinfo {year} {2014})},\ \Eprint {http://arxiv.org/abs/1201.2400}
  {arXiv:1201.2400 [hep-ph]} \BibitemShut {NoStop}%
\bibitem [{\citenamefont {Kouvaris}\ and\ \citenamefont
  {Tinyakov}(2014)}]{Kouvaris:2013kra}%
  \BibitemOpen
  \bibfield  {author} {\bibinfo {author} {\bibfnamefont {Chris}\ \bibnamefont
  {Kouvaris}}\ and\ \bibinfo {author} {\bibfnamefont {Peter}\ \bibnamefont
  {Tinyakov}},\ }\bibfield  {title} {\enquote {\bibinfo {title} {{Growth of
  Black Holes in the interior of Rotating Neutron Stars}},}\ }\href {\doibase
  10.1103/PhysRevD.90.043512} {\bibfield  {journal} {\bibinfo  {journal} {Phys.
  Rev.}\ }\textbf {\bibinfo {volume} {D90}},\ \bibinfo {pages} {043512}
  (\bibinfo {year} {2014})},\ \Eprint {http://arxiv.org/abs/1312.3764}
  {arXiv:1312.3764 [astro-ph.SR]} \BibitemShut {NoStop}%
\bibitem [{\citenamefont {Kurita}\ and\ \citenamefont
  {Nakano}(2016)}]{Kurita:2015vga}%
  \BibitemOpen
  \bibfield  {author} {\bibinfo {author} {\bibfnamefont {Yasunari}\
  \bibnamefont {Kurita}}\ and\ \bibinfo {author} {\bibfnamefont {Hiroyuki}\
  \bibnamefont {Nakano}},\ }\bibfield  {title} {\enquote {\bibinfo {title}
  {{Gravitational waves from dark matter collapse in a star}},}\ }\href
  {\doibase 10.1103/PhysRevD.93.023508} {\bibfield  {journal} {\bibinfo
  {journal} {Phys. Rev.}\ }\textbf {\bibinfo {volume} {D93}},\ \bibinfo {pages}
  {023508} (\bibinfo {year} {2016})},\ \Eprint
  {http://arxiv.org/abs/1510.00893} {arXiv:1510.00893 [gr-qc]} \BibitemShut
  {NoStop}%
\bibitem [{\citenamefont {Kouvaris}(2012)}]{Kouvaris:2011gb}%
  \BibitemOpen
  \bibfield  {author} {\bibinfo {author} {\bibfnamefont {Chris}\ \bibnamefont
  {Kouvaris}},\ }\bibfield  {title} {\enquote {\bibinfo {title} {{Limits on
  Self-Interacting Dark Matter}},}\ }\href {\doibase
  10.1103/PhysRevLett.108.191301} {\bibfield  {journal} {\bibinfo  {journal}
  {Phys. Rev. Lett.}\ }\textbf {\bibinfo {volume} {108}},\ \bibinfo {pages}
  {191301} (\bibinfo {year} {2012})},\ \Eprint {http://arxiv.org/abs/1111.4364}
  {arXiv:1111.4364 [astro-ph.CO]} \BibitemShut {NoStop}%
\bibitem [{\citenamefont {Bramante}\ \emph {et~al.}(2014)\citenamefont
  {Bramante}, \citenamefont {Fukushima}, \citenamefont {Kumar},\ and\
  \citenamefont {Stopnitzky}}]{Bramante:2013nma}%
  \BibitemOpen
  \bibfield  {author} {\bibinfo {author} {\bibfnamefont {Joseph}\ \bibnamefont
  {Bramante}}, \bibinfo {author} {\bibfnamefont {Keita}\ \bibnamefont
  {Fukushima}}, \bibinfo {author} {\bibfnamefont {Jason}\ \bibnamefont
  {Kumar}}, \ and\ \bibinfo {author} {\bibfnamefont {Elan}\ \bibnamefont
  {Stopnitzky}},\ }\bibfield  {title} {\enquote {\bibinfo {title} {{Bounds on
  self-interacting fermion dark matter from observations of old neutron
  stars}},}\ }\href {\doibase 10.1103/PhysRevD.89.015010} {\bibfield  {journal}
  {\bibinfo  {journal} {Phys. Rev.}\ }\textbf {\bibinfo {volume} {D89}},\
  \bibinfo {pages} {015010} (\bibinfo {year} {2014})},\ \Eprint
  {http://arxiv.org/abs/1310.3509} {arXiv:1310.3509 [hep-ph]} \BibitemShut
  {NoStop}%
\bibitem [{\citenamefont {Bramante}(2015)}]{Bramante:2015cua}%
  \BibitemOpen
  \bibfield  {author} {\bibinfo {author} {\bibfnamefont {Joseph}\ \bibnamefont
  {Bramante}},\ }\bibfield  {title} {\enquote {\bibinfo {title} {{Dark matter
  ignition of type Ia supernovae}},}\ }\href {\doibase
  10.1103/PhysRevLett.115.141301} {\bibfield  {journal} {\bibinfo  {journal}
  {Phys. Rev. Lett.}\ }\textbf {\bibinfo {volume} {115}},\ \bibinfo {pages}
  {141301} (\bibinfo {year} {2015})},\ \Eprint
  {http://arxiv.org/abs/1505.07464} {arXiv:1505.07464 [hep-ph]} \BibitemShut
  {NoStop}%
\bibitem [{\citenamefont {Bramante}\ and\ \citenamefont
  {Unwin}(2017)}]{Bramante:2017obj}%
  \BibitemOpen
  \bibfield  {author} {\bibinfo {author} {\bibfnamefont {Joseph}\ \bibnamefont
  {Bramante}}\ and\ \bibinfo {author} {\bibfnamefont {James}\ \bibnamefont
  {Unwin}},\ }\bibfield  {title} {\enquote {\bibinfo {title} {{Superheavy
  Thermal Dark Matter and Primordial Asymmetries}},}\ }\href {\doibase
  10.1007/JHEP02(2017)119} {\bibfield  {journal} {\bibinfo  {journal} {JHEP}\
  }\textbf {\bibinfo {volume} {02}},\ \bibinfo {pages} {119} (\bibinfo {year}
  {2017})},\ \Eprint {http://arxiv.org/abs/1701.05859} {arXiv:1701.05859
  [hep-ph]} \BibitemShut {NoStop}%
\bibitem [{\citenamefont {Baryakhtar}\ \emph {et~al.}(2017)\citenamefont
  {Baryakhtar}, \citenamefont {Bramante}, \citenamefont {Li}, \citenamefont
  {Linden},\ and\ \citenamefont {Raj}}]{Baryakhtar:2017dbj}%
  \BibitemOpen
  \bibfield  {author} {\bibinfo {author} {\bibfnamefont {Masha}\ \bibnamefont
  {Baryakhtar}}, \bibinfo {author} {\bibfnamefont {Joseph}\ \bibnamefont
  {Bramante}}, \bibinfo {author} {\bibfnamefont {Shirley~Weishi}\ \bibnamefont
  {Li}}, \bibinfo {author} {\bibfnamefont {Tim}\ \bibnamefont {Linden}}, \ and\
  \bibinfo {author} {\bibfnamefont {Nirmal}\ \bibnamefont {Raj}},\ }\bibfield
  {title} {\enquote {\bibinfo {title} {{Dark Kinetic Heating of Neutron Stars
  and An Infrared Window On WIMPs, SIMPs, and Pure Higgsinos}},}\ }\href@noop
  {} {\  (\bibinfo {year} {2017})},\ \Eprint {http://arxiv.org/abs/1704.01577}
  {arXiv:1704.01577 [hep-ph]} \BibitemShut {NoStop}%
\bibitem [{\citenamefont {Colpi}\ \emph {et~al.}(1986)\citenamefont {Colpi},
  \citenamefont {Shapiro},\ and\ \citenamefont {Wasserman}}]{Colpi:1986ye}%
  \BibitemOpen
  \bibfield  {author} {\bibinfo {author} {\bibfnamefont {M.}~\bibnamefont
  {Colpi}}, \bibinfo {author} {\bibfnamefont {S.~L.}\ \bibnamefont {Shapiro}},
  \ and\ \bibinfo {author} {\bibfnamefont {I.}~\bibnamefont {Wasserman}},\
  }\bibfield  {title} {\enquote {\bibinfo {title} {{Boson Stars: Gravitational
  Equilibria of Selfinteracting Scalar Fields}},}\ }\href {\doibase
  10.1103/PhysRevLett.57.2485} {\bibfield  {journal} {\bibinfo  {journal}
  {Phys. Rev. Lett.}\ }\textbf {\bibinfo {volume} {57}},\ \bibinfo {pages}
  {2485--2488} (\bibinfo {year} {1986})}\BibitemShut {NoStop}%
\bibitem [{\citenamefont {{Lattimer}}\ \emph {et~al.}(1977)\citenamefont
  {{Lattimer}}, \citenamefont {{Mackie}}, \citenamefont {{Ravenhall}},\ and\
  \citenamefont {{Schramm}}}]{1977ApJ...213..225L}%
  \BibitemOpen
  \bibfield  {author} {\bibinfo {author} {\bibfnamefont {J.~M.}\ \bibnamefont
  {{Lattimer}}}, \bibinfo {author} {\bibfnamefont {F.}~\bibnamefont
  {{Mackie}}}, \bibinfo {author} {\bibfnamefont {D.~G.}\ \bibnamefont
  {{Ravenhall}}}, \ and\ \bibinfo {author} {\bibfnamefont {D.~N.}\ \bibnamefont
  {{Schramm}}},\ }\bibfield  {title} {\enquote {\bibinfo {title} {{The
  decompression of cold neutron star matter}},}\ }\href {\doibase
  10.1086/155148} {\bibfield  {journal} {\bibinfo  {journal} {\apj}\ }\textbf
  {\bibinfo {volume} {213}},\ \bibinfo {pages} {225--233} (\bibinfo {year}
  {1977})}\BibitemShut {NoStop}%
\bibitem [{\citenamefont {Kasen}\ \emph {et~al.}(2013)\citenamefont {Kasen},
  \citenamefont {Badnell},\ and\ \citenamefont {Barnes}}]{Kasen:2013xka}%
  \BibitemOpen
  \bibfield  {author} {\bibinfo {author} {\bibfnamefont {Daniel}\ \bibnamefont
  {Kasen}}, \bibinfo {author} {\bibfnamefont {N.~R.}\ \bibnamefont {Badnell}},
  \ and\ \bibinfo {author} {\bibfnamefont {Jennifer}\ \bibnamefont {Barnes}},\
  }\bibfield  {title} {\enquote {\bibinfo {title} {{Opacities and Spectra of
  the $r$-process Ejecta from Neutron Star Mergers}},}\ }\href {\doibase
  10.1088/0004-637X/774/1/25} {\bibfield  {journal} {\bibinfo  {journal}
  {Astrophys. J.}\ }\textbf {\bibinfo {volume} {774}},\ \bibinfo {pages} {25}
  (\bibinfo {year} {2013})},\ \Eprint {http://arxiv.org/abs/1303.5788}
  {arXiv:1303.5788 [astro-ph.HE]} \BibitemShut {NoStop}%
\bibitem [{\citenamefont {Abadie}\ \emph {et~al.}(2010)\citenamefont {Abadie}
  \emph {et~al.}}]{Abadie:2010cf}%
  \BibitemOpen
  \bibfield  {author} {\bibinfo {author} {\bibfnamefont {J.}~\bibnamefont
  {Abadie}} \emph {et~al.} (\bibinfo {collaboration} {VIRGO, LIGO
  Scientific}),\ }\bibfield  {title} {\enquote {\bibinfo {title} {{Predictions
  for the Rates of Compact Binary Coalescences Observable by Ground-based
  Gravitational-wave Detectors}},}\ }\href {\doibase
  10.1088/0264-9381/27/17/173001} {\bibfield  {journal} {\bibinfo  {journal}
  {Class. Quant. Grav.}\ }\textbf {\bibinfo {volume} {27}},\ \bibinfo {pages}
  {173001} (\bibinfo {year} {2010})},\ \Eprint {http://arxiv.org/abs/1003.2480}
  {arXiv:1003.2480 [astro-ph.HE]} \BibitemShut {NoStop}%
\bibitem [{\citenamefont {Hopkins}\ and\ \citenamefont
  {Beacom}(2006)}]{Hopkins:2006bw}%
  \BibitemOpen
  \bibfield  {author} {\bibinfo {author} {\bibfnamefont {Andrew~M.}\
  \bibnamefont {Hopkins}}\ and\ \bibinfo {author} {\bibfnamefont {John~F.}\
  \bibnamefont {Beacom}},\ }\bibfield  {title} {\enquote {\bibinfo {title} {{On
  the normalisation of the cosmic star formation history}},}\ }\href {\doibase
  10.1086/506610} {\bibfield  {journal} {\bibinfo  {journal} {Astrophys. J.}\
  }\textbf {\bibinfo {volume} {651}},\ \bibinfo {pages} {142--154} (\bibinfo
  {year} {2006})},\ \Eprint {http://arxiv.org/abs/astro-ph/0601463}
  {arXiv:astro-ph/0601463 [astro-ph]} \BibitemShut {NoStop}%
\bibitem [{\citenamefont {{Paczynski}}(1990)}]{1990ApJ...348..485P}%
  \BibitemOpen
  \bibfield  {author} {\bibinfo {author} {\bibfnamefont {B.}~\bibnamefont
  {{Paczynski}}},\ }\bibfield  {title} {\enquote {\bibinfo {title} {{A test of
  the galactic origin of gamma-ray bursts}},}\ }\href {\doibase 10.1086/168257}
  {\bibfield  {journal} {\bibinfo  {journal} {\apj}\ }\textbf {\bibinfo
  {volume} {348}},\ \bibinfo {pages} {485--494} (\bibinfo {year}
  {1990})}\BibitemShut {NoStop}%
\bibitem [{\citenamefont {{Blaes}}\ and\ \citenamefont
  {{Madau}}(1993)}]{1993ApJ...403..690B}%
  \BibitemOpen
  \bibfield  {author} {\bibinfo {author} {\bibfnamefont {O.}~\bibnamefont
  {{Blaes}}}\ and\ \bibinfo {author} {\bibfnamefont {P.}~\bibnamefont
  {{Madau}}},\ }\bibfield  {title} {\enquote {\bibinfo {title} {{Can we observe
  accreting, isolated neutron stars?}}}\ }\href {\doibase 10.1086/172240}
  {\bibfield  {journal} {\bibinfo  {journal} {\apj}\ }\textbf {\bibinfo
  {volume} {403}},\ \bibinfo {pages} {690--705} (\bibinfo {year}
  {1993})}\BibitemShut {NoStop}%
\bibitem [{\citenamefont {{Sartore}}\ \emph {et~al.}(2010)\citenamefont
  {{Sartore}}, \citenamefont {{Ripamonti}}, \citenamefont {{Treves}},\ and\
  \citenamefont {{Turolla}}}]{2010A&A...510A..23S}%
  \BibitemOpen
  \bibfield  {author} {\bibinfo {author} {\bibfnamefont {N.}~\bibnamefont
  {{Sartore}}}, \bibinfo {author} {\bibfnamefont {E.}~\bibnamefont
  {{Ripamonti}}}, \bibinfo {author} {\bibfnamefont {A.}~\bibnamefont
  {{Treves}}}, \ and\ \bibinfo {author} {\bibfnamefont {R.}~\bibnamefont
  {{Turolla}}},\ }\bibfield  {title} {\enquote {\bibinfo {title} {{Galactic
  neutron stars. I. Space and velocity distributions in the disk and in the
  halo}},}\ }\href {\doibase 10.1051/0004-6361/200912222} {\ \textbf {\bibinfo
  {volume} {510}},\ \bibinfo {eid} {A23} (\bibinfo {year} {2010})},\ \Eprint
  {http://arxiv.org/abs/0908.3182} {arXiv:0908.3182} \BibitemShut {NoStop}%
\bibitem [{\citenamefont {Hobbs}\ \emph {et~al.}(2005)\citenamefont {Hobbs},
  \citenamefont {Lorimer}, \citenamefont {Lyne},\ and\ \citenamefont
  {Kramer}}]{Hobbs:2005yx}%
  \BibitemOpen
  \bibfield  {author} {\bibinfo {author} {\bibfnamefont {George}\ \bibnamefont
  {Hobbs}}, \bibinfo {author} {\bibfnamefont {D.~R.}\ \bibnamefont {Lorimer}},
  \bibinfo {author} {\bibfnamefont {A.~G.}\ \bibnamefont {Lyne}}, \ and\
  \bibinfo {author} {\bibfnamefont {M.}~\bibnamefont {Kramer}},\ }\bibfield
  {title} {\enquote {\bibinfo {title} {{A Statistical study of 233 pulsar
  proper motions}},}\ }\href {\doibase 10.1111/j.1365-2966.2005.09087.x}
  {\bibfield  {journal} {\bibinfo  {journal} {Mon. Not. Roy. Astron. Soc.}\
  }\textbf {\bibinfo {volume} {360}},\ \bibinfo {pages} {974--992} (\bibinfo
  {year} {2005})},\ \Eprint {http://arxiv.org/abs/astro-ph/0504584}
  {arXiv:astro-ph/0504584 [astro-ph]} \BibitemShut {NoStop}%
\bibitem [{\citenamefont {Faucher-Giguere}\ and\ \citenamefont
  {Kaspi}(2006)}]{FaucherGiguere:2005ny}%
  \BibitemOpen
  \bibfield  {author} {\bibinfo {author} {\bibfnamefont {Claude-Andre}\
  \bibnamefont {Faucher-Giguere}}\ and\ \bibinfo {author} {\bibfnamefont
  {Victoria~M.}\ \bibnamefont {Kaspi}},\ }\bibfield  {title} {\enquote
  {\bibinfo {title} {{Birth and evolution of isolated radio pulsars}},}\ }\href
  {\doibase 10.1086/501516} {\bibfield  {journal} {\bibinfo  {journal}
  {Astrophys. J.}\ }\textbf {\bibinfo {volume} {643}},\ \bibinfo {pages}
  {332--355} (\bibinfo {year} {2006})},\ \Eprint
  {http://arxiv.org/abs/astro-ph/0512585} {arXiv:astro-ph/0512585 [astro-ph]}
  \BibitemShut {NoStop}%
\bibitem [{\citenamefont {Navarro}\ \emph {et~al.}(1997)\citenamefont
  {Navarro}, \citenamefont {Frenk},\ and\ \citenamefont
  {White}}]{Navarro:1996gj}%
  \BibitemOpen
  \bibfield  {author} {\bibinfo {author} {\bibfnamefont {Julio~F.}\
  \bibnamefont {Navarro}}, \bibinfo {author} {\bibfnamefont {Carlos~S.}\
  \bibnamefont {Frenk}}, \ and\ \bibinfo {author} {\bibfnamefont {Simon D.~M.}\
  \bibnamefont {White}},\ }\bibfield  {title} {\enquote {\bibinfo {title} {{A
  Universal density profile from hierarchical clustering}},}\ }\href {\doibase
  10.1086/304888} {\bibfield  {journal} {\bibinfo  {journal} {Astrophys. J.}\
  }\textbf {\bibinfo {volume} {490}},\ \bibinfo {pages} {493--508} (\bibinfo
  {year} {1997})},\ \Eprint {http://arxiv.org/abs/astro-ph/9611107}
  {arXiv:astro-ph/9611107 [astro-ph]} \BibitemShut {NoStop}%
\bibitem [{\citenamefont {Sofue}(2013)}]{Sofue:2013kja}%
  \BibitemOpen
  \bibfield  {author} {\bibinfo {author} {\bibfnamefont {Yoshiaki}\
  \bibnamefont {Sofue}},\ }\bibfield  {title} {\enquote {\bibinfo {title}
  {{Rotation Curve and Mass Distribution in the Galactic Center --- From Black
  Hole to Entire Galaxy ---}},}\ }\href {\doibase 10.1093/pasj/65.6.118}
  {\bibfield  {journal} {\bibinfo  {journal} {Publ. Astron. Soc. Jap.}\
  }\textbf {\bibinfo {volume} {65}},\ \bibinfo {pages} {118} (\bibinfo {year}
  {2013})},\ \Eprint {http://arxiv.org/abs/1307.8241} {arXiv:1307.8241
  [astro-ph.GA]} \BibitemShut {NoStop}%
\bibitem [{\citenamefont {Hawking}(1974)}]{Hawking:1974rv}%
  \BibitemOpen
  \bibfield  {author} {\bibinfo {author} {\bibfnamefont {S.~W.}\ \bibnamefont
  {Hawking}},\ }\bibfield  {title} {\enquote {\bibinfo {title} {{Black hole
  explosions}},}\ }\href {\doibase 10.1038/248030a0} {\bibfield  {journal}
  {\bibinfo  {journal} {Nature}\ }\textbf {\bibinfo {volume} {248}},\ \bibinfo
  {pages} {30--31} (\bibinfo {year} {1974})}\BibitemShut {NoStop}%
\bibitem [{\citenamefont {Carr}\ and\ \citenamefont
  {Hawking}(1974)}]{Carr:1974nx}%
  \BibitemOpen
  \bibfield  {author} {\bibinfo {author} {\bibfnamefont {Bernard~J.}\
  \bibnamefont {Carr}}\ and\ \bibinfo {author} {\bibfnamefont {S.~W.}\
  \bibnamefont {Hawking}},\ }\bibfield  {title} {\enquote {\bibinfo {title}
  {{Black holes in the early Universe}},}\ }\href@noop {} {\bibfield  {journal}
  {\bibinfo  {journal} {Mon. Not. Roy. Astron. Soc.}\ }\textbf {\bibinfo
  {volume} {168}},\ \bibinfo {pages} {399--415} (\bibinfo {year}
  {1974})}\BibitemShut {NoStop}%
\bibitem [{\citenamefont {Capela}\ \emph
  {et~al.}(2013{\natexlab{a}})\citenamefont {Capela}, \citenamefont
  {Pshirkov},\ and\ \citenamefont {Tinyakov}}]{Capela:2013yf}%
  \BibitemOpen
  \bibfield  {author} {\bibinfo {author} {\bibfnamefont {Fabio}\ \bibnamefont
  {Capela}}, \bibinfo {author} {\bibfnamefont {Maxim}\ \bibnamefont
  {Pshirkov}}, \ and\ \bibinfo {author} {\bibfnamefont {Peter}\ \bibnamefont
  {Tinyakov}},\ }\bibfield  {title} {\enquote {\bibinfo {title} {{Constraints
  on primordial black holes as dark matter candidates from capture by neutron
  stars}},}\ }\href {\doibase 10.1103/PhysRevD.87.123524} {\bibfield  {journal}
  {\bibinfo  {journal} {Phys. Rev.}\ }\textbf {\bibinfo {volume} {D87}},\
  \bibinfo {pages} {123524} (\bibinfo {year} {2013}{\natexlab{a}})},\ \Eprint
  {http://arxiv.org/abs/1301.4984} {arXiv:1301.4984 [astro-ph.CO]} \BibitemShut
  {NoStop}%
\bibitem [{\citenamefont {Capela}\ \emph
  {et~al.}(2013{\natexlab{b}})\citenamefont {Capela}, \citenamefont
  {Pshirkov},\ and\ \citenamefont {Tinyakov}}]{Capela:2012jz}%
  \BibitemOpen
  \bibfield  {author} {\bibinfo {author} {\bibfnamefont {Fabio}\ \bibnamefont
  {Capela}}, \bibinfo {author} {\bibfnamefont {Maxim}\ \bibnamefont
  {Pshirkov}}, \ and\ \bibinfo {author} {\bibfnamefont {Peter}\ \bibnamefont
  {Tinyakov}},\ }\bibfield  {title} {\enquote {\bibinfo {title} {{Constraints
  on Primordial Black Holes as Dark Matter Candidates from Star Formation}},}\
  }\href {\doibase 10.1103/PhysRevD.87.023507} {\bibfield  {journal} {\bibinfo
  {journal} {Phys. Rev.}\ }\textbf {\bibinfo {volume} {D87}},\ \bibinfo {pages}
  {023507} (\bibinfo {year} {2013}{\natexlab{b}})},\ \Eprint
  {http://arxiv.org/abs/1209.6021} {arXiv:1209.6021 [astro-ph.CO]} \BibitemShut
  {NoStop}%
\bibitem [{\citenamefont {Fuller}\ \emph {et~al.}(2017)\citenamefont {Fuller},
  \citenamefont {Kusenko},\ and\ \citenamefont {Takhistov}}]{Fuller:2017uyd}%
  \BibitemOpen
  \bibfield  {author} {\bibinfo {author} {\bibfnamefont {George~M.}\
  \bibnamefont {Fuller}}, \bibinfo {author} {\bibfnamefont {Alexander}\
  \bibnamefont {Kusenko}}, \ and\ \bibinfo {author} {\bibfnamefont {Volodymyr}\
  \bibnamefont {Takhistov}},\ }\bibfield  {title} {\enquote {\bibinfo {title}
  {{Primordial Black Holes and $r$-Process Nucleosynthesis}},}\ }\href@noop {}
  {\  (\bibinfo {year} {2017})},\ \Eprint {http://arxiv.org/abs/1704.01129}
  {arXiv:1704.01129 [astro-ph.HE]} \BibitemShut {NoStop}%
\bibitem [{\citenamefont {{Eichler}}\ \emph {et~al.}(1989)\citenamefont
  {{Eichler}}, \citenamefont {{Livio}}, \citenamefont {{Piran}},\ and\
  \citenamefont {{Schramm}}}]{1989Natur.340..126E}%
  \BibitemOpen
  \bibfield  {author} {\bibinfo {author} {\bibfnamefont {D.}~\bibnamefont
  {{Eichler}}}, \bibinfo {author} {\bibfnamefont {M.}~\bibnamefont {{Livio}}},
  \bibinfo {author} {\bibfnamefont {T.}~\bibnamefont {{Piran}}}, \ and\
  \bibinfo {author} {\bibfnamefont {D.~N.}\ \bibnamefont {{Schramm}}},\
  }\bibfield  {title} {\enquote {\bibinfo {title} {{Nucleosynthesis, neutrino
  bursts and gamma-rays from coalescing neutron stars}},}\ }\href {\doibase
  10.1038/340126a0} {\bibfield  {journal} {\bibinfo  {journal} {\nat}\ }\textbf
  {\bibinfo {volume} {340}},\ \bibinfo {pages} {126--128} (\bibinfo {year}
  {1989})}\BibitemShut {NoStop}%
\bibitem [{\citenamefont {{Ji}}\ \emph {et~al.}(2016)\citenamefont {{Ji}},
  \citenamefont {{Frebel}}, \citenamefont {{Chiti}},\ and\ \citenamefont
  {{Simon}}}]{2016Natur.531..610J}%
  \BibitemOpen
  \bibfield  {author} {\bibinfo {author} {\bibfnamefont {A.~P.}\ \bibnamefont
  {{Ji}}}, \bibinfo {author} {\bibfnamefont {A.}~\bibnamefont {{Frebel}}},
  \bibinfo {author} {\bibfnamefont {A.}~\bibnamefont {{Chiti}}}, \ and\
  \bibinfo {author} {\bibfnamefont {J.~D.}\ \bibnamefont {{Simon}}},\
  }\bibfield  {title} {\enquote {\bibinfo {title} {{R-process enrichment from a
  single event in an ancient dwarf galaxy}},}\ }\href {\doibase
  10.1038/nature17425} {\bibfield  {journal} {\bibinfo  {journal} {\nat}\
  }\textbf {\bibinfo {volume} {531}},\ \bibinfo {pages} {610--613} (\bibinfo
  {year} {2016})},\ \Eprint {http://arxiv.org/abs/1512.01558}
  {arXiv:1512.01558} \BibitemShut {NoStop}%
\bibitem [{\citenamefont {{van de Voort}}\ \emph {et~al.}(2015)\citenamefont
  {{van de Voort}}, \citenamefont {{Quataert}}, \citenamefont {{Hopkins}},
  \citenamefont {{Kere{\v s}}},\ and\ \citenamefont
  {{Faucher-Gigu{\`e}re}}}]{2015MNRAS.447..140V}%
  \BibitemOpen
  \bibfield  {author} {\bibinfo {author} {\bibfnamefont {F.}~\bibnamefont {{van
  de Voort}}}, \bibinfo {author} {\bibfnamefont {E.}~\bibnamefont
  {{Quataert}}}, \bibinfo {author} {\bibfnamefont {P.~F.}\ \bibnamefont
  {{Hopkins}}}, \bibinfo {author} {\bibfnamefont {D.}~\bibnamefont {{Kere{\v
  s}}}}, \ and\ \bibinfo {author} {\bibfnamefont {C.-A.}\ \bibnamefont
  {{Faucher-Gigu{\`e}re}}},\ }\bibfield  {title} {\enquote {\bibinfo {title}
  {{Galactic r-process enrichment by neutron star mergers in cosmological
  simulations of a Milky Way-mass galaxy}},}\ }\href {\doibase
  10.1093/mnras/stu2404} {\ \textbf {\bibinfo {volume} {447}},\ \bibinfo
  {pages} {140--148} (\bibinfo {year} {2015})},\ \Eprint
  {http://arxiv.org/abs/1407.7039} {arXiv:1407.7039} \BibitemShut {NoStop}%
\bibitem [{\citenamefont {{Shen}}\ \emph {et~al.}(2015)\citenamefont {{Shen}},
  \citenamefont {{Cooke}}, \citenamefont {{Ramirez-Ruiz}}, \citenamefont
  {{Madau}}, \citenamefont {{Mayer}},\ and\ \citenamefont
  {{Guedes}}}]{2015ApJ...807..115S}%
  \BibitemOpen
  \bibfield  {author} {\bibinfo {author} {\bibfnamefont {S.}~\bibnamefont
  {{Shen}}}, \bibinfo {author} {\bibfnamefont {R.~J.}\ \bibnamefont {{Cooke}}},
  \bibinfo {author} {\bibfnamefont {E.}~\bibnamefont {{Ramirez-Ruiz}}},
  \bibinfo {author} {\bibfnamefont {P.}~\bibnamefont {{Madau}}}, \bibinfo
  {author} {\bibfnamefont {L.}~\bibnamefont {{Mayer}}}, \ and\ \bibinfo
  {author} {\bibfnamefont {J.}~\bibnamefont {{Guedes}}},\ }\bibfield  {title}
  {\enquote {\bibinfo {title} {{The History of R-Process Enrichment in the
  Milky Way}},}\ }\href {\doibase 10.1088/0004-637X/807/2/115} {\bibfield
  {journal} {\bibinfo  {journal} {\apj}\ }\textbf {\bibinfo {volume} {807}},\
  \bibinfo {eid} {115} (\bibinfo {year} {2015})},\ \Eprint
  {http://arxiv.org/abs/1407.3796} {arXiv:1407.3796} \BibitemShut {NoStop}%
\bibitem [{\citenamefont {Wehmeyer}\ \emph {et~al.}(2015)\citenamefont
  {Wehmeyer}, \citenamefont {Pignatari},\ and\ \citenamefont
  {Thielemann}}]{Wehmeyer:2015sra}%
  \BibitemOpen
  \bibfield  {author} {\bibinfo {author} {\bibfnamefont {B.}~\bibnamefont
  {Wehmeyer}}, \bibinfo {author} {\bibfnamefont {M.}~\bibnamefont {Pignatari}},
  \ and\ \bibinfo {author} {\bibfnamefont {F.~K.}\ \bibnamefont {Thielemann}},\
  }\bibfield  {title} {\enquote {\bibinfo {title} {{Galactic evolution of rapid
  neutron capture process abundances: the inhomogeneous approach}},}\ }\href
  {\doibase 10.1093/mnras/stv1352} {\bibfield  {journal} {\bibinfo  {journal}
  {Mon. Not. Roy. Astron. Soc.}\ }\textbf {\bibinfo {volume} {452}},\ \bibinfo
  {pages} {1970--1981} (\bibinfo {year} {2015})},\ \Eprint
  {http://arxiv.org/abs/1501.07749} {arXiv:1501.07749 [astro-ph.GA]}
  \BibitemShut {NoStop}%
\bibitem [{\citenamefont {Aprile}\ \emph {et~al.}(2017)\citenamefont {Aprile}
  \emph {et~al.}}]{Aprile:2017iyp}%
  \BibitemOpen
  \bibfield  {author} {\bibinfo {author} {\bibfnamefont {E.}~\bibnamefont
  {Aprile}} \emph {et~al.} (\bibinfo {collaboration} {XENON}),\ }\bibfield
  {title} {\enquote {\bibinfo {title} {{First Dark Matter Search Results from
  the XENON1T Experiment}},}\ }\href {\doibase 10.1103/PhysRevLett.119.181301}
  {\bibfield  {journal} {\bibinfo  {journal} {Phys. Rev. Lett.}\ }\textbf
  {\bibinfo {volume} {119}},\ \bibinfo {pages} {181301} (\bibinfo {year}
  {2017})},\ \Eprint {http://arxiv.org/abs/1705.06655} {arXiv:1705.06655
  [astro-ph.CO]} \BibitemShut {NoStop}%
\bibitem [{\citenamefont {Ruppin}\ \emph {et~al.}(2014)\citenamefont {Ruppin},
  \citenamefont {Billard}, \citenamefont {Figueroa-Feliciano},\ and\
  \citenamefont {Strigari}}]{Ruppin:2014bra}%
  \BibitemOpen
  \bibfield  {author} {\bibinfo {author} {\bibfnamefont {F.}~\bibnamefont
  {Ruppin}}, \bibinfo {author} {\bibfnamefont {J.}~\bibnamefont {Billard}},
  \bibinfo {author} {\bibfnamefont {E.}~\bibnamefont {Figueroa-Feliciano}}, \
  and\ \bibinfo {author} {\bibfnamefont {L.}~\bibnamefont {Strigari}},\
  }\bibfield  {title} {\enquote {\bibinfo {title} {{Complementarity of dark
  matter detectors in light of the neutrino background}},}\ }\href {\doibase
  10.1103/PhysRevD.90.083510} {\bibfield  {journal} {\bibinfo  {journal} {Phys.
  Rev.}\ }\textbf {\bibinfo {volume} {D90}},\ \bibinfo {pages} {083510}
  (\bibinfo {year} {2014})},\ \Eprint {http://arxiv.org/abs/1408.3581}
  {arXiv:1408.3581 [hep-ph]} \BibitemShut {NoStop}%
\bibitem [{\citenamefont {{Bellm}}(2014)}]{2014htu..conf...27B}%
  \BibitemOpen
  \bibfield  {author} {\bibinfo {author} {\bibfnamefont {E.}~\bibnamefont
  {{Bellm}}},\ }\bibfield  {title} {\enquote {\bibinfo {title} {{The Zwicky
  Transient Facility}},}\ }in\ \href@noop {} {\emph {\bibinfo {booktitle} {The
  Third Hot-wiring the Transient Universe Workshop}}},\ \bibinfo {editor}
  {edited by\ \bibinfo {editor} {\bibfnamefont {P.~R.}\ \bibnamefont
  {{Wozniak}}}, \bibinfo {editor} {\bibfnamefont {M.~J.}\ \bibnamefont
  {{Graham}}}, \bibinfo {editor} {\bibfnamefont {A.~A.}\ \bibnamefont
  {{Mahabal}}}, \ and\ \bibinfo {editor} {\bibfnamefont {R.}~\bibnamefont
  {{Seaman}}}}\ (\bibinfo {year} {2014})\ pp.\ \bibinfo {pages} {27--33},\
  \Eprint {http://arxiv.org/abs/1410.8185} {arXiv:1410.8185 [astro-ph.IM]}
  \BibitemShut {NoStop}%
\bibitem [{\citenamefont {Blanchard}\ \emph {et~al.}(2017)\citenamefont
  {Blanchard} \emph {et~al.}}]{Blanchard:2017csd}%
  \BibitemOpen
  \bibfield  {author} {\bibinfo {author} {\bibfnamefont {P.~K.}\ \bibnamefont
  {Blanchard}} \emph {et~al.},\ }\bibfield  {title} {\enquote {\bibinfo {title}
  {{The Electromagnetic Counterpart of the Binary Neutron Star Merger
  LIGO/VIRGO GW170817. VII. Properties of the Host Galaxy and Constraints on
  the Merger Timescale}},}\ }\href {\doibase 10.3847/2041-8213/aa9055}
  {\bibfield  {journal} {\bibinfo  {journal} {Astrophys. J.}\ }\textbf
  {\bibinfo {volume} {848}},\ \bibinfo {pages} {L22} (\bibinfo {year}
  {2017})},\ \Eprint {http://arxiv.org/abs/1710.05458} {arXiv:1710.05458
  [astro-ph.HE]} \BibitemShut {NoStop}%
\bibitem [{\citenamefont {Bramante}\ \emph {et~al.}(2017)\citenamefont
  {Bramante}, \citenamefont {Delgado},\ and\ \citenamefont
  {Martin}}]{Bramante:2017xlb}%
  \BibitemOpen
  \bibfield  {author} {\bibinfo {author} {\bibfnamefont {Joseph}\ \bibnamefont
  {Bramante}}, \bibinfo {author} {\bibfnamefont {Antonio}\ \bibnamefont
  {Delgado}}, \ and\ \bibinfo {author} {\bibfnamefont {Adam}\ \bibnamefont
  {Martin}},\ }\bibfield  {title} {\enquote {\bibinfo {title} {{Multiscatter
  stellar capture of dark matter}},}\ }\href {\doibase
  10.1103/PhysRevD.96.063002} {\bibfield  {journal} {\bibinfo  {journal} {Phys.
  Rev.}\ }\textbf {\bibinfo {volume} {D96}},\ \bibinfo {pages} {063002}
  (\bibinfo {year} {2017})},\ \Eprint {http://arxiv.org/abs/1703.04043}
  {arXiv:1703.04043 [hep-ph]} \BibitemShut {NoStop}%
\bibitem [{\citenamefont {Rane}\ \emph {et~al.}(2016)\citenamefont {Rane},
  \citenamefont {Lorimer}, \citenamefont {Bates}, \citenamefont {McMann},
  \citenamefont {McLaughlin},\ and\ \citenamefont {Rajwade}}]{Rane:2015sxa}%
  \BibitemOpen
  \bibfield  {author} {\bibinfo {author} {\bibfnamefont {A.}~\bibnamefont
  {Rane}}, \bibinfo {author} {\bibfnamefont {D.~R.}\ \bibnamefont {Lorimer}},
  \bibinfo {author} {\bibfnamefont {S.~D.}\ \bibnamefont {Bates}}, \bibinfo
  {author} {\bibfnamefont {N.}~\bibnamefont {McMann}}, \bibinfo {author}
  {\bibfnamefont {M.~A.}\ \bibnamefont {McLaughlin}}, \ and\ \bibinfo {author}
  {\bibfnamefont {K.}~\bibnamefont {Rajwade}},\ }\bibfield  {title} {\enquote
  {\bibinfo {title} {{A search for rotating radio transients and fast radio
  bursts in the Parkes high-latitude pulsar survey}},}\ }\href {\doibase
  10.1093/mnras/stv2404} {\bibfield  {journal} {\bibinfo  {journal} {Mon. Not.
  Roy. Astron. Soc.}\ }\textbf {\bibinfo {volume} {455}},\ \bibinfo {pages}
  {2207--2215} (\bibinfo {year} {2016})},\ \Eprint
  {http://arxiv.org/abs/1505.00834} {arXiv:1505.00834 [astro-ph.HE]}
  \BibitemShut {NoStop}%
\bibitem [{\citenamefont {Wiel}\ \emph {et~al.}(2016)\citenamefont {Wiel},
  \citenamefont {Burke-Spolaor}, \citenamefont {Lawrence}, \citenamefont
  {Law},\ and\ \citenamefont {Bower}}]{Wiel:2016pdl}%
  \BibitemOpen
  \bibfield  {author} {\bibinfo {author} {\bibfnamefont {Scott~Vander}\
  \bibnamefont {Wiel}}, \bibinfo {author} {\bibfnamefont {Sarah}\ \bibnamefont
  {Burke-Spolaor}}, \bibinfo {author} {\bibfnamefont {Earl}\ \bibnamefont
  {Lawrence}}, \bibinfo {author} {\bibfnamefont {Casey~J.}\ \bibnamefont
  {Law}}, \ and\ \bibinfo {author} {\bibfnamefont {Geoffrey~C.}\ \bibnamefont
  {Bower}},\ }\bibfield  {title} {\enquote {\bibinfo {title} {{Rare Event
  Statistics Applied to Fast Radio Bursts}},}\ }\href@noop {} {\  (\bibinfo
  {year} {2016})},\ \Eprint {http://arxiv.org/abs/1612.00896} {arXiv:1612.00896
  [astro-ph.IM]} \BibitemShut {NoStop}%
\bibitem [{\citenamefont {Lorimer}\ \emph {et~al.}(2007)\citenamefont
  {Lorimer}, \citenamefont {Bailes}, \citenamefont {McLaughlin}, \citenamefont
  {Narkevic},\ and\ \citenamefont {Crawford}}]{Lorimer:2007qn}%
  \BibitemOpen
  \bibfield  {author} {\bibinfo {author} {\bibfnamefont {D.~R.}\ \bibnamefont
  {Lorimer}}, \bibinfo {author} {\bibfnamefont {M.}~\bibnamefont {Bailes}},
  \bibinfo {author} {\bibfnamefont {M.~A.}\ \bibnamefont {McLaughlin}},
  \bibinfo {author} {\bibfnamefont {D.~J.}\ \bibnamefont {Narkevic}}, \ and\
  \bibinfo {author} {\bibfnamefont {F.}~\bibnamefont {Crawford}},\ }\bibfield
  {title} {\enquote {\bibinfo {title} {{A bright millisecond radio burst of
  extragalactic origin}},}\ }\href {\doibase 10.1126/science.1147532}
  {\bibfield  {journal} {\bibinfo  {journal} {Science}\ }\textbf {\bibinfo
  {volume} {318}},\ \bibinfo {pages} {777} (\bibinfo {year} {2007})},\ \Eprint
  {http://arxiv.org/abs/0709.4301} {arXiv:0709.4301 [astro-ph]} \BibitemShut
  {NoStop}%
\bibitem [{\citenamefont {Thornton}\ \emph {et~al.}(2013)\citenamefont
  {Thornton} \emph {et~al.}}]{Thornton:2013iua}%
  \BibitemOpen
  \bibfield  {author} {\bibinfo {author} {\bibfnamefont {D.}~\bibnamefont
  {Thornton}} \emph {et~al.},\ }\bibfield  {title} {\enquote {\bibinfo {title}
  {{A Population of Fast Radio Bursts at Cosmological Distances}},}\ }\href
  {\doibase 10.1126/science.1236789} {\bibfield  {journal} {\bibinfo  {journal}
  {Science}\ }\textbf {\bibinfo {volume} {341}},\ \bibinfo {pages} {53--56}
  (\bibinfo {year} {2013})},\ \Eprint {http://arxiv.org/abs/1307.1628}
  {arXiv:1307.1628 [astro-ph.HE]} \BibitemShut {NoStop}%
\bibitem [{\citenamefont {Palenzuela}(2013)}]{Palenzuela:2012my}%
  \BibitemOpen
  \bibfield  {author} {\bibinfo {author} {\bibfnamefont {Carlos}\ \bibnamefont
  {Palenzuela}},\ }\bibfield  {title} {\enquote {\bibinfo {title} {{Modeling
  magnetized neutron stars using resistive MHD}},}\ }\href {\doibase
  10.1093/mnras/stt311} {\bibfield  {journal} {\bibinfo  {journal} {Mon. Not.
  Roy. Astron. Soc.}\ }\textbf {\bibinfo {volume} {431}},\ \bibinfo {pages}
  {1853--1865} (\bibinfo {year} {2013})},\ \Eprint
  {http://arxiv.org/abs/1212.0130} {arXiv:1212.0130 [astro-ph.HE]} \BibitemShut
  {NoStop}%
\bibitem [{\citenamefont {Dionysopoulou}\ \emph {et~al.}(2013)\citenamefont
  {Dionysopoulou}, \citenamefont {Alic}, \citenamefont {Palenzuela},
  \citenamefont {Rezzolla},\ and\ \citenamefont
  {Giacomazzo}}]{Dionysopoulou:2012zv}%
  \BibitemOpen
  \bibfield  {author} {\bibinfo {author} {\bibfnamefont {Kyriaki}\ \bibnamefont
  {Dionysopoulou}}, \bibinfo {author} {\bibfnamefont {Daniela}\ \bibnamefont
  {Alic}}, \bibinfo {author} {\bibfnamefont {Carlos}\ \bibnamefont
  {Palenzuela}}, \bibinfo {author} {\bibfnamefont {Luciano}\ \bibnamefont
  {Rezzolla}}, \ and\ \bibinfo {author} {\bibfnamefont {Bruno}\ \bibnamefont
  {Giacomazzo}},\ }\bibfield  {title} {\enquote {\bibinfo {title}
  {{General-Relativistic Resistive Magnetohydrodynamics in three dimensions:
  formulation and tests}},}\ }\href {\doibase 10.1103/PhysRevD.88.044020}
  {\bibfield  {journal} {\bibinfo  {journal} {Phys. Rev.}\ }\textbf {\bibinfo
  {volume} {D88}},\ \bibinfo {pages} {044020} (\bibinfo {year} {2013})},\
  \Eprint {http://arxiv.org/abs/1208.3487} {arXiv:1208.3487 [gr-qc]}
  \BibitemShut {NoStop}%
\bibitem [{\citenamefont {Falcke}\ and\ \citenamefont
  {Rezzolla}(2014)}]{Falcke:2013xpa}%
  \BibitemOpen
  \bibfield  {author} {\bibinfo {author} {\bibfnamefont {Heino}\ \bibnamefont
  {Falcke}}\ and\ \bibinfo {author} {\bibfnamefont {Luciano}\ \bibnamefont
  {Rezzolla}},\ }\bibfield  {title} {\enquote {\bibinfo {title} {{Fast radio
  bursts: the last sign of supramassive neutron stars}},}\ }\href {\doibase
  10.1051/0004-6361/201321996} {\bibfield  {journal} {\bibinfo  {journal}
  {Astron. Astrophys.}\ }\textbf {\bibinfo {volume} {562}},\ \bibinfo {pages}
  {A137} (\bibinfo {year} {2014})},\ \Eprint {http://arxiv.org/abs/1307.1409}
  {arXiv:1307.1409 [astro-ph.HE]} \BibitemShut {NoStop}%
\bibitem [{\citenamefont {Diehl}\ \emph {et~al.}(2006)\citenamefont {Diehl}
  \emph {et~al.}}]{Diehl:2006cf}%
  \BibitemOpen
  \bibfield  {author} {\bibinfo {author} {\bibfnamefont {Roland}\ \bibnamefont
  {Diehl}} \emph {et~al.},\ }\bibfield  {title} {\enquote {\bibinfo {title}
  {{Radioactive Al-26 and massive stars in the galaxy}},}\ }\href {\doibase
  10.1038/nature04364} {\bibfield  {journal} {\bibinfo  {journal} {Nature}\
  }\textbf {\bibinfo {volume} {439}},\ \bibinfo {pages} {45--47} (\bibinfo
  {year} {2006})},\ \Eprint {http://arxiv.org/abs/astro-ph/0601015}
  {arXiv:astro-ph/0601015 [astro-ph]} \BibitemShut {NoStop}%
\bibitem [{\citenamefont {{Robitaille}}\ and\ \citenamefont
  {{Whitney}}(2010)}]{2010ApJ...710L..11R}%
  \BibitemOpen
  \bibfield  {author} {\bibinfo {author} {\bibfnamefont {T.~P.}\ \bibnamefont
  {{Robitaille}}}\ and\ \bibinfo {author} {\bibfnamefont {B.~A.}\ \bibnamefont
  {{Whitney}}},\ }\bibfield  {title} {\enquote {\bibinfo {title} {{The
  Present-Day Star Formation Rate of the Milky Way Determined from
  Spitzer-Detected Young Stellar Objects}},}\ }\href {\doibase
  10.1088/2041-8205/710/1/L11} {\bibfield  {journal} {\bibinfo  {journal}
  {ApJL}\ }\textbf {\bibinfo {volume} {710}},\ \bibinfo {pages} {L11--L15}
  (\bibinfo {year} {2010})},\ \Eprint {http://arxiv.org/abs/1001.3672}
  {arXiv:1001.3672 [astro-ph.GA]} \BibitemShut {NoStop}%
\bibitem [{\citenamefont {van Dokkum}\ \emph {et~al.}(2013)\citenamefont {van
  Dokkum} \emph {et~al.}}]{vanDokkum:2013hza}%
  \BibitemOpen
  \bibfield  {author} {\bibinfo {author} {\bibfnamefont {Pieter~G.}\
  \bibnamefont {van Dokkum}} \emph {et~al.},\ }\bibfield  {title} {\enquote
  {\bibinfo {title} {{The Assembly of Milky Way-like Galaxies Since z~2.5}},}\
  }\href {\doibase 10.1088/2041-8205/771/2/L35} {\bibfield  {journal} {\bibinfo
   {journal} {Astrophys. J.}\ }\textbf {\bibinfo {volume} {771}},\ \bibinfo
  {pages} {L35} (\bibinfo {year} {2013})},\ \Eprint
  {http://arxiv.org/abs/1304.2391} {arXiv:1304.2391 [astro-ph.CO]} \BibitemShut
  {NoStop}%
\bibitem [{\citenamefont {{Snaith}}\ \emph {et~al.}(2014)\citenamefont
  {{Snaith}}, \citenamefont {{Haywood}}, \citenamefont {{Di Matteo}},
  \citenamefont {{Lehnert}}, \citenamefont {{Combes}}, \citenamefont {{Katz}},\
  and\ \citenamefont {{G{\'o}mez}}}]{2014ApJ...781L..31S}%
  \BibitemOpen
  \bibfield  {author} {\bibinfo {author} {\bibfnamefont {O.~N.}\ \bibnamefont
  {{Snaith}}}, \bibinfo {author} {\bibfnamefont {M.}~\bibnamefont {{Haywood}}},
  \bibinfo {author} {\bibfnamefont {P.}~\bibnamefont {{Di Matteo}}}, \bibinfo
  {author} {\bibfnamefont {M.~D.}\ \bibnamefont {{Lehnert}}}, \bibinfo {author}
  {\bibfnamefont {F.}~\bibnamefont {{Combes}}}, \bibinfo {author}
  {\bibfnamefont {D.}~\bibnamefont {{Katz}}}, \ and\ \bibinfo {author}
  {\bibfnamefont {A.}~\bibnamefont {{G{\'o}mez}}},\ }\bibfield  {title}
  {\enquote {\bibinfo {title} {{The Dominant Epoch of Star Formation in the
  Milky Way Formed the Thick Disk}},}\ }\href {\doibase
  10.1088/2041-8205/781/2/L31} {\bibfield  {journal} {\bibinfo  {journal}
  {ApJL}\ }\textbf {\bibinfo {volume} {781}},\ \bibinfo {eid} {L31} (\bibinfo
  {year} {2014})},\ \Eprint {http://arxiv.org/abs/1401.1835} {arXiv:1401.1835}
  \BibitemShut {NoStop}%
\bibitem [{\citenamefont {{Marasco}}\ \emph {et~al.}(2015)\citenamefont
  {{Marasco}}, \citenamefont {{Debattista}}, \citenamefont {{Fraternali}},
  \citenamefont {{van der Hulst}}, \citenamefont {{Wadsley}}, \citenamefont
  {{Quinn}},\ and\ \citenamefont {{Ro{\v s}kar}}}]{2015MNRAS.451.4223M}%
  \BibitemOpen
  \bibfield  {author} {\bibinfo {author} {\bibfnamefont {A.}~\bibnamefont
  {{Marasco}}}, \bibinfo {author} {\bibfnamefont {V.~P.}\ \bibnamefont
  {{Debattista}}}, \bibinfo {author} {\bibfnamefont {F.}~\bibnamefont
  {{Fraternali}}}, \bibinfo {author} {\bibfnamefont {T.}~\bibnamefont {{van der
  Hulst}}}, \bibinfo {author} {\bibfnamefont {J.}~\bibnamefont {{Wadsley}}},
  \bibinfo {author} {\bibfnamefont {T.}~\bibnamefont {{Quinn}}}, \ and\
  \bibinfo {author} {\bibfnamefont {R.}~\bibnamefont {{Ro{\v s}kar}}},\
  }\bibfield  {title} {\enquote {\bibinfo {title} {{The effect of stellar
  feedback on a Milky Way-like galaxy and its gaseous halo}},}\ }\href
  {\doibase 10.1093/mnras/stv1240} {\bibfield  {journal} {\bibinfo  {journal}
  {MNRAS}\ }\textbf {\bibinfo {volume} {451}},\ \bibinfo {pages} {4223--4237}
  (\bibinfo {year} {2015})},\ \Eprint {http://arxiv.org/abs/1506.00652}
  {arXiv:1506.00652} \BibitemShut {NoStop}%
\bibitem [{\citenamefont {Baiotti}\ \emph {et~al.}(2007)\citenamefont
  {Baiotti}, \citenamefont {Hawke},\ and\ \citenamefont
  {Rezzolla}}]{Baiotti:2007np}%
  \BibitemOpen
  \bibfield  {author} {\bibinfo {author} {\bibfnamefont {Luca}\ \bibnamefont
  {Baiotti}}, \bibinfo {author} {\bibfnamefont {Ian}\ \bibnamefont {Hawke}}, \
  and\ \bibinfo {author} {\bibfnamefont {Luciano}\ \bibnamefont {Rezzolla}},\
  }\bibfield  {title} {\enquote {\bibinfo {title} {{On the Gravitational
  radiation from the collapse of neutron stars to rotating black holes}},}\
  }\href {\doibase 10.1088/0264-9381/24/12/S13} {\bibfield  {journal} {\bibinfo
   {journal} {Class. Quant. Grav.}\ }\textbf {\bibinfo {volume} {24}},\
  \bibinfo {pages} {S187--S206} (\bibinfo {year} {2007})},\ \Eprint
  {http://arxiv.org/abs/gr-qc/0701043} {arXiv:gr-qc/0701043 [gr-qc]}
  \BibitemShut {NoStop}%
\bibitem [{\citenamefont {Abbott}\ \emph {et~al.}(2013)\citenamefont {Abbott}
  \emph {et~al.}}]{Aasi:2013wya}%
  \BibitemOpen
  \bibfield  {author} {\bibinfo {author} {\bibfnamefont {B.~P.}\ \bibnamefont
  {Abbott}} \emph {et~al.} (\bibinfo {collaboration} {VIRGO, LIGO
  Scientific}),\ }\bibfield  {title} {\enquote {\bibinfo {title} {{Prospects
  for Observing and Localizing Gravitational-Wave Transients with Advanced LIGO
  and Advanced Virgo}},}\ }\href {\doibase 10.1007/lrr-2016-1} {\  (\bibinfo
  {year} {2013}),\ 10.1007/lrr-2016-1},\ \bibinfo {note} {[Living Rev.
  Rel.19,1(2016)]},\ \Eprint {http://arxiv.org/abs/1304.0670} {arXiv:1304.0670
  [gr-qc]} \BibitemShut {NoStop}%
\bibitem [{\citenamefont {Giudice}\ \emph {et~al.}(2016)\citenamefont
  {Giudice}, \citenamefont {McCullough},\ and\ \citenamefont
  {Urbano}}]{Giudice:2016zpa}%
  \BibitemOpen
  \bibfield  {author} {\bibinfo {author} {\bibfnamefont {Gian~F.}\ \bibnamefont
  {Giudice}}, \bibinfo {author} {\bibfnamefont {Matthew}\ \bibnamefont
  {McCullough}}, \ and\ \bibinfo {author} {\bibfnamefont {Alfredo}\
  \bibnamefont {Urbano}},\ }\bibfield  {title} {\enquote {\bibinfo {title}
  {{Hunting for Dark Particles with Gravitational Waves}},}\ }\href {\doibase
  10.1088/1475-7516/2016/10/001} {\bibfield  {journal} {\bibinfo  {journal}
  {JCAP}\ }\textbf {\bibinfo {volume} {1610}},\ \bibinfo {pages} {001}
  (\bibinfo {year} {2016})},\ \Eprint {http://arxiv.org/abs/1605.01209}
  {arXiv:1605.01209 [hep-ph]} \BibitemShut {NoStop}%
\end{thebibliography}%

\end{document}